\documentclass[letterpaper,12pt]{article}%
\usepackage{setspace}
\usepackage{amsfonts}
\usepackage{amssymb, amsmath, amsthm,  graphicx}
\usepackage[margin=1in]{geometry}
\usepackage[bottom]{footmisc}
\usepackage{commath}
\usepackage[utf8]{inputenc}
\usepackage{natbib}
\usepackage{booktabs, multirow}
\usepackage{url}
\usepackage{amsmath}
\usepackage{amssymb}
\usepackage{graphicx}%
\providecommand{\U}[1]{\protect\rule{.1in}{.1in}}

\newtheorem{thm}{Theorem}
\newtheorem{definition}{Definition}

\newtheorem{lemma}{Lemma}

\newtheorem{assum}{Assumption}
\begin{document}

\title{Should Humans Lie to Machines? \\The Incentive Compatibility of Lasso and General Weighted Lasso\thanks{We thank Anders Kock, Jos${\Acute{e}}$ Luis Montiel Olea, Ran Spiegler
and seminar participants at Simon Fraser University for their valuable
comments. We are grateful for the hospitality of the Economics Department at
Columbia University, where this research is initiated when both authors were
visitors in 2018-2019. Eliaz gratefully acknowledges financial support from
ISF grant 470/19. }}
\author{\textsc{Mehmet Caner\thanks{North Carolina State University, Nelson Hall,
Department of Economics, NC 27695. Email:mcaner@ncsu.edu. } }
\and \textsc{Kfir Eliaz\thanks{School of Economics, Tel-Aviv University and Eccles
School of Business, the University of Utah. Email:\ kfire@tauex.tau.ac.il.}} }
\date{\today}
\maketitle

\begin{abstract}
We consider situations where a user feeds her attributes to a machine learning
method that tries to predict her best option based on a random sample of other
users. The predictor is incentive-compatible if the user has no incentive to
misreport her covariates. Focusing on the popular Lasso estimation technique,
we borrow tools from high-dimensional statistics to characterize sufficient
conditions that ensure that Lasso is incentive compatible in large samples. We
extend our results to the Conservative Lasso estimator and provide new moment
bounds for this generalized weighted version of Lasso. Our results show that
incentive compatibility is achieved if the tuning parameter is kept above some
threshold. We present simulations that illustrate how this can be done in practice.

\end{abstract}

\section{Introduction}

Rapid advances in machine learning methods for analyzing big data have given
rise to automated systems that employ these methods to predict the best
fitting outcomes for users based on their personal characteristics. For
example, many online platforms try to predict which content - a song, a video,
a post, or an article - is the best fit for each user. Medical providers have
also begun using machine learning techniques to automate check-ups and test
appointments for patients based on their medical history. Typically, these
automated systems use data from past users to estimate a model that relates
the best fit for a user (such as the most preferred content or the appropriate
medical test) to her characteristics. These estimates are then applied to a
new user's characteristics, which she discloses either actively or passively
via her past online behavior (which may be reflected in her cookies or
collected by her browser). Given the growing interaction of users with such
automated systems, it is only natural to ask whether a user should truthfully
disclose her characteristics?

If the information the user discloses is also used to exploit her (say, by
providing it to third parties for advertising or price discrimination), then
the user has an obvious reason not to reveal her private information. The
question is whether special features of some popular machine learning methods
introduce an incentive to misreport one's personal characteristics even when
this information will be used \textit{solely} for predicting her best
outcome?\footnote{In a recent interview of Brian Christian, the author of
\textit{The Alignment Problem}, he notes that \textquotedblleft computers may
one day be able not only to learn our behavior but also intuit our values -
figure out from our actions what it is we're trying to optimize. ... What if
an algorithm intuits the `wrong' values, based on its best read of who we
currently are but not of who we aspire to be? Do we really want our computers
inferring our values form browser histories? See \cite{shaywitz20} for this
interview.} This question is of crucial importance:\ If individuals submit
false reports to systems that rely on these reports for estimation and
predictions, then the conclusions drawn from such estimates and predictions
will be wrong and may lead to quite undesirable outcomes (e.g., think of an
automated medical platform that schedules tests for patients based on false
reports on attributes such as smoking, drinking and physical exercise).

To address the above question, we consider a stylized environment where each
user $i$'s ideal option is a linear function $f$ of her privately observed
attributes $X_{i}=(X_{i,1},...,X_{i,p})^{\prime}$ such that $f(X_{i}%
)=X_{i}^{\prime}\mathbf{\beta_{0}}.$ A user may not know the values of the
coefficients $\mathbf{\beta_{0},}$ in which case she would have some (possibly
degenerate) prior beliefs over them. A \textquotedblleft
statistician\textquotedblright, who represents some automated prediction
platform has a sample of the attributes of $n$ users and \textit{noisy}
observations on their ideal options. For instance, suppose $f(X_{i})$ is the
optimal dosage of some medication when taken immediately at the onset of
symptoms, conditional on the patient's medical history $X_{i},$ but the
statistician observes the dosage that was given after some delay. Similarly,
$f(X_{i})$ may be the mix of news and reality shows that a user with
attributes $X_{i}$ actually watches, but the statistician observes only self
reports by a user who may have forgotten exactly what he watched.

The statistician uses her sample to estimate the function $f$ by computing an
estimate $\mathbf{\hat{\beta}}$ of the true coefficients $\mathbf{\beta_{0}}$.
The statistician wishes to apply these estimates to predict the ideal option
of a new user, $n+1,$ whose true attributes $X_{n+1}$ are not observed by the
statistician. This new user must decide what vector of attributes
$R(X_{n+1})\ $(which may \textit{differ} from the truth) to report to the
statistician. 

In making this decision, the new user takes into account her beliefs about the
statistician's sample (the new user only knows the distribution from which the
sample is drawn, but she does not observe its realization), and her beliefs
about the true parameters $\beta_{0}$. The statistician then plugs the new
user's reported attributes into the estimated function and gives the user the
option $R(X_{n+1})^{\prime}\hat{\beta},$ which is the statistician's estimate
of the user's ideal option based her report. The new user's expected loss from
a report $R(X_{n+1})$ is given by the mean square error between her
expectation of the ideal option $X_{n+1}^{\prime}\beta_{0}$ and her assigned
option $R(X_{n+1})^{\prime}\hat{\beta}.$ The statistician's estimator is
\textit{incentive-compatible}, if the new user has no incentive to deviate
from truthful reporting whatever her attributes are, and for \textit{any}
prior belief on $\beta_{0}:$ I.e., if for every possible value of $\beta_{0}$
and $X_{n+1},$ the expected value of $(X_{n+1}^{\prime}\beta_{0}%
-R(X_{n+1})^{\prime}\hat{\beta})^{2}$ is minimized at the truth $R(X_{n+1}%
)=X_{n+1},$ where the expectation is taken with respect to the statistician's sample.

Intuition suggests that an individual cannot benefit from lying to a procedure
that is meant to predict the best outcome for her. To counter this intuition,
\cite{es2019}, and \cite{es2020} use the above framework to illustrate that a
user may have a strict incentive to lie about her attributes when the
prediction is based on a linear regression that penalizes non-zero estimated
coefficients. The rough intuition is that the user believes that despite the
statistician's good intentions, these estimation techniques lead to
distortions, which she tries to undo by lying. For instance, given the user's
beliefs about the true model parameters, she may be concerned that the
estimator will admit too many irrelevant attributes, and hence, she reports a
zero value for these attributes (see \cite{es2019}, and \cite{es2020} for more
details). However, these papers focus on particular examples in which
attributes are \textit{binary}, the statistician has the \textit{same} (fixed)
finite number of observations on each possible combination of attribute
values, and the penalty parameter is \textit{fixed} and does \textit{not}
adjust to the sample size. That is, these papers only raise the problem of
incentive compatibility but do not provide an econometric solution. Hence,
they leave open the following important question: For a general environment,
are there conditions ensuring that a penalized regression model is incentive
compatible\textit{ }in large samples?

Answering this question can potentially allow platforms, like those discussed
above, to use machine-learning methods to predict users' most preferred
options without worrying that their data is \textquotedblleft
contaminated\textquotedblright\ by non-truthful users. Put bluntly, estimates
and predictions made by methods that are \textit{not} incentive-compatible are
possibly unreliable since they may be based on false data.

This paper addresses the above open question by first focusing on the most
popular form of penalized regressions - the \textit{Lasso}
estimator.\footnote{Our results can be extended to apply to the debiased lasso
estimator, but this involves a different proof technique, and hence, is beyond
the scope of the current paper.} Borrowing tools from high-dimensional
statistics, we establish sufficient conditions for incentive compatibility of
the Lasso estimator in large samples. We show that to achieve incentive
compatibility, the tuning parameter must be \textit{large} enough (i.e., it
must remain above some threshold as sample size increases) so as to avoid
overfitting, which is the main reason why a user may want to lie (see Remark 2
in Section 4). This potential to lie implies that the standard way of choosing
small enough tuning parameters to ensure consistency may violate incentive
compatibility. We provide simulation results that illustrate how the tuning
parameter can be chosen in practice to ensure incentive compatibility.
Incentive compatibility may therefore be viewed as an additional important
property that should be imposed on estimators on top of consistency and unbiasedness.

Next, we extend our results to a general weighted Lasso, also known as the
\textquotedblleft Conservative Lasso\textquotedblright. \cite{ck18} develop
this estimator as a data-dependent weighted penalized estimator. Conservative
Lasso better differentiates between relevant and irrelevant variables, which
results in better $l_{2}$ norm errors. The superior model selection properties
of the Conservative Lasso (compared to the standard Lasso) is shown in
\cite{ck18} analytically as well as in simulations. We characterize the
conditions for ensuring the incentive-compatibility of the Conservative Lasso
in large samples, and show this may require a higher (relative to the standard
Lasso) lower bound for the tuning parameter under certain scenarios.

We also offer a new technical contribution by extending the
oracle-moment-inequalities of \cite{jvdg18} from sub-Gaussian to i.i.d. data.
Using a different proof technique, we derive less conservative bounds on the
moments of the Lasso estimator and relax the bounded signal to noise ratio
assumption in \cite{jvdg18}. We also extend \cite{jvdg18} from Lasso moment
estimation to generalized weighted Lasso (Conservative Lasso). It is shown
that moment bound estimation results cover also this general class of penalty.
These are all new results for general weighted Lasso.


The motivation to focus first on the Lasso estimator stems from the fact that
this estimator is the benchmark among all high dimensional statistical
estimators that predict large scale models when the number of regressors
exceeds the sample size. Following its original proposal by
\cite{tibshirani96}, econometricians and statisticians have used Lasso-based
estimators to push the boundaries of economics and finance. One of the most
critical issues facing these Lasso type estimators is post-inference after
estimation and model selection, which require uniformly valid confidence
intervals. In a seminal series of papers, Belloni et al. (2012,2014) solved
these issues by introducing the idea of \textquotedblleft partialling
out\textquotedblright\ the regressors. A different, but complementary
approach, via debiasing-desparsifying is proposed by \cite{van2014}.
\cite{ck18} extended the debiasing of \cite{van2014} to
heteroskedastic-non-sub-Gaussian data with strong oracle optimality
properties, thereby proposing a high dimensional estimator that is robust to
heteroskedasticity, and with uniformly valid confidence intervals. Lasso-based
debiasing are used in panel data models (see, e.g., \cite{cgst18}, \cite{k16},
\cite{kt19}) and for addressing quantile treatment effects and text analysis
(see, e.g., \cite{cs19} and \cite{c20}).


The concern that statistical procedures such as estimation, forecasting and
classification are vulnerable to manipulation, has been the subject of some
recent papers in the computer science literature. In contrast to us, this
literature assumes there is an explicit conflict of interest between the
statistician and the data providers - either because the latter are concerned
about their privacy, they have to incur a cost to provide a precise report, or
they have a different objective than the statistician. These papers analyze
the Nash equilibria of a game where users submit private values that are used
for estimation/classification, and propose incentive schemes that induce
truthful reporting. Some notable works in this literature include
\cite{cdp15}, \cite{cil15}, \cite{dfp10}, \cite{gvz15}, \cite{hmpw16},
\cite{mpr12} and \cite{pp04}. \textit{None} of these papers consider penalized
regression methods, and \textit{none} of them characterize conditions
guaranteeing incentive compatibility of regression techniques when the
statistician and users have \textit{aligned interests} (as is the case in our model).

The remainder of the paper is organized as follows. Section 2 introduces our
model and assumptions. Section 3 provides new oracle inequalities. Section 4
characterizes the sufficient conditions for ensuring that Lasso is incentive
compatible in large samples. Section 5 extends these results to general
weighted Lasso. Section 6 provides simulation results and Section 7 concludes.
Appendix A contains the proofs of the results on the Lasso estimator when the
number of regressors ($p$) exceed the number of observations ($n$). Appendix B
addresses the case of $p\leq n$ and shows how to extend our Lasso resuls when
we relax our assumption on the signal to noise ratio. Finally, Appendix C
contains the proofs for the general weighted Lasso.

\section{ The model}

We begin this section by describing our theoretical framework and introducing
our notion of incentive-compatibility. We then discuss the key ingredients of
our model and conclude by laying out our assumptions on the statistician's data.

Throughout the paper we will use the following notational conventions. For any
vector $\nu\in R^{d}$, let $\Vert\nu\Vert_{1},\Vert\nu\Vert_{2},\Vert\nu
\Vert_{\infty}$ denote its $l_{1},l_{2},l_{\infty}$ norm respectively, and
$\Vert\nu\Vert_{0}$ be the $l_{0}$ norm, which means the total number of
nonzero entries. For a set $S\subseteq\{1,2,\cdots,d\}$, let $|S|=s$ be the
cardinality of the set. Let $\nu_{S}$ be the modified $\nu$ such that we put 0
when the index does not belong to $S$ (i.e., say $S=\{1,2,6\}$ for a
$10\times1$ vector $\nu$, this means that $\nu$ is modified such that now all
elements are zero except elements $1,2,6$). Let $\Vert A\Vert_{l_{1}}$ be the
maximum absolute column-sum norm of a matrix of dimensions $m\times l$, i.e.,
$\Vert A\Vert_{l_{1}}=\max_{1\leq k\leq l}\sum_{i=1}^{m}|A_{ik}|$ which is
also called the induced $l_{1}$ norm of $A$. Let $\| A \|_{l_{\infty}}:=
\max_{1 \le i \le m} \sum_{k=1}^{l} |A_{ik}|$ which is the maximum absolute row
sum norm.

Our environment consists of users who are characterized by a set of $p$
personal characteristics. For instance, in the context of medical decision
making, a characteristic can represent a risk factor (obesity, smoking, etc.).
For each user $i,$ these characteristics are modeled as $p$ explanatory
variables, $X_{i,1},...,X_{i,p},$ drawn from some distribution over a subset
of $%
\mathbb{R}
^{p}$. These attributes determine the ideal option for a user according to the
function%
\[
f(X_{i,1},...,X_{i,p})=\sum_{k=1}^{p}X_{i,k}\beta_{0,k}%
\]
This function applies to all users, who differ only in the values of their
characteristics. The realized values of $(X_{i,1},...,X_{i,p})$ are privately
observed by user $i.$ A user may or may not know the value of the coefficients
$(\beta_{0,1},...,\beta_{0,p}).$ In the latter case, she has some (possibly
degenerate) prior beliefs over their values.

A \textit{statistician} (representing the automated prediction systems
described in the introduction) has\textit{ private} access to a sample of $n$
observations. Each observation $i=1,...,n$ consists of the true attributes
$X_{i}=(X_{i,1},...,X_{i,p})$ of user $i$ and a noisy signal $y_{i}$ of that
user's ideal option,
\begin{equation}
y_{i}=X_{i}^{\prime}\beta_{0}+u_{i}, \label{model}%
\end{equation}
where $u_{i}$ is random noise that is drawn $i.i.d$ from some distribution
with zero mean.\footnote{Access to such observations is a necessary condition
for any platform that tries to learn about users (say, Netflix, Spotify). In
the introduction, we gave a couple of examples for such data, which may be
obtained from a third party, or from marketing surveys.}

The $X_{i}$' s are also i.i.d. across $i$ and exogenous, and will be discussed
in detail in Assumption \ref{a1} in the next subsection. $\beta_{0}$ is a
$p\times1$ vector, representing the true parameters in $f$. We let
$S_{0}=\{j:\beta_{0,j}\neq0\}$ denote the set of relevant regressors with
$s_{0}$ being the cardinality of the set $S_{0}$. (i.e., $s_{0}$ of the
elements of $\beta_{0}$ are nonzero, and the rest are zero). $s_{0}$ is a
nondecreasing function of $n$, and we assume $s_{0}\geq1$. These facts are
known to an \textquotedblleft oracle\textquotedblright\ but not to the
statistician (and possibly not to a user).

\subsection{The Lasso Estimator}

Using her (privately observed) sample, the statistician estimates the function
$f$, or equivalently, she estimates the coefficients $\beta_{0,1}%
,...,\beta_{0,p}$. When $p>n$, the least squares estimator is infeasible due
to singularity of the empirical Gram matrix. Hence, the statistician uses
Lasso, the penalized regression procedure that assigns costs to including
explanatory variables in the regression. Specifically, she solves the
following minimization problem%
\begin{equation}
\hat{\beta}=argmin_{\beta\in R^{p}}\frac{\sum_{i=1}^{n}(y_{i}-X_{i}^{\prime
}\beta)^{2}}{n}+2\lambda_{n}\Vert\beta\Vert_{1}, \label{lasso}%
\end{equation}
where $\lambda_{n}>0$ is the penalty (also called tuning parameter)\ that
decreases with the number of observations at the rate of $\lambda_{n}%
=O(\sqrt{lnp/n})$ (an explicit expression for the sequence $\lambda_{n}$ is
given in equation (\ref{lambda}) in Appendix A).\footnote{We established this
rate in Lemma \ref{l2} in Appendix A.}

Given her estimates $\hat{\beta},$ the statistician must take an action $a\in%
\mathbb{R}
$ on behalf of a \textit{new} user, $j=n+1$. This action is just the
statistician's prediction of the ideal option of that user. The new user's
payoff from action $a$ is $-(a-f(X_{n+1}))^{2}$, where $f(X_{n+1})$ is the
true ideal option associated with her personal attributes $X_{n+1}$.

Since the statistician does not observe $X_{n+1},$ in order to make her
prediction of $f(X_{n+1}),$ she asks the $n+1$ user to report a $p\times1$
vector, $R (X_{n+1}),$ which is interpreted as that user's attributes. The
statistician then plugs $R (X_{n+1})$ into her estimated model and chooses the
action $a=R (X_{n+1})^{\prime}\hat{\beta}.$ When the $n+1$ user decides what
attribute values to report, she takes into account that she does not observe
the statistician's sample, and hence, does not know the values of the
estimated coefficients $\hat{\beta}.$ She only knows the distribution from
which the statistician's sample is drawn, and that given her sample, the
statistician chooses $\hat{\beta}$ according to (\ref{lasso}). Given this, the
user chooses the report $R (X_{n+1})$ that minimize her expected loss
$E_{\beta_{0},\hat{\beta}}(R (X_{n+1})^{\prime}\hat{\beta}-X_{n+1}^{\prime
}\beta_{0})^{2},$ where the expectation is taken with respect to the user's
prior beliefs about the true parameters $\beta_{0},$ and her beliefs about the
estimate $\hat{\beta}.$ Hence, the new user may decide to lie and report $R
(X_{n+1})\neq X_{n+1}.$ In particular, she may decide to \textquotedblleft opt
out\textquotedblright\ and submit a vector of zeros.\footnote{In the case in
which the individual's attributes are collected \textquotedblleft
passively\textquotedblright\ from her browsing history, then reporting a
vector of zero attributes can be interpreted as the act of deleting cookies.}
Our objective is to understand under what conditions it is in the user's best
interest to be truthful regardless of her prior beliefs on $\beta_{0}.$

\subsection{Incentive Compatibility}

To introduce our notion of \textit{incentive compatibility, }consider a user
who upon observing her vector of covariates decides which vector of values to
report (which may differ from the true values). An estimator is said to be
(ex-post) incentive-compatible, if for \textit{any} vector of covariates, and
for \textit{any} belief over the true model parameters, the user's expected
payoff from truthful reporting is at least as high as her expected payoff from
any misreport, where the expectation is taken with respect to the
statistician's sample.

\begin{definition}
\label{ic} An estimator is \textbf{incentive-compatible} if for every
$X_{n+1},$ for every $R(X_{n+1})$ and for every every $\beta_{0},$
\begin{equation}
E[R(X_{n+1})^{\prime}\hat{\beta}-X_{n+1}^{\prime}\beta_{0}]^{2}\geq
E[X_{n+1}^{\prime}\hat{\beta}-X_{n+1}^{\prime}\beta_{0}]^{2}. \label{IC}%
\end{equation}
where the expectation $E$ is taken with respect to the possible realizations
of the statistician's sample.
\end{definition}

Incentive compatibility means that the user is unable to perform better by
misreporting her personal characteristics, \textit{regardless} of her beliefs
over the true model's parameters in mean squared sense.\footnote{If we were to
relax the requirement that truth-telling is preferred for \textit{every} prior
belief over the true model's parameters, we would need to make some
assumptions on the user's prior beliefs (see, e.g. Eliaz and Spiegler (2020)).
Thus, our incentive-compatibility has the merit of being robust to
\textit{any} specification of prior beliefs.} How should we interpret this
requirement, given that we do not necessarily want to think of the user as
being sophisticated enough to think in these terms? One interpretation is that
lack of incentive compatibility is merely a \textit{normative} statement about
the user's welfare - namely, given our model of how the statistician takes
actions on the user's behalf, it would be advisable for her to misrepresent
her personal characteristics. Furthermore, there are opportunities for new
firms to enter and offer the user paid advice for how to manipulate the
procedure - in analogy to the industry of \textquotedblleft search engine
optimization\textquotedblright. Incentive compatibility theoretically
eliminates the need for such an industry. In the context of the online content
provision story, some misreporting strategies take the form of
\textquotedblleft deleting cookies\textquotedblright. This deviation is
straightforward to implement, and the user can check if it makes her better
off in the long run.

Note that incentive-compatibility is not a property that can be tested
statistically. To see this, suppose each user is characterized by only a
single covariate that is uniformly distributed on $\{0,1\}.$ If users are
truthful, then one would expect a 50-50 distribution of $0$'s and $1$'s in the
population. However, if each user lies about his covariate, then one would
also observe a 50-50 distribution of $0$'s and $1$'s.

Recall that the statistician's sample contains the \textit{true} attributes of
$n$ users. The idea is that the data on these users is obtained through a
different process than the way the statistician obtains the data from the
$n+1$ user. For instance, as mentioned earlier, this data may be obtained from
a marketing survey where there is no incentive to lie. Alternatively, one may
interpret our incentive compatibility requirement as a requirement that
truth-telling is a \textit{Nash equilibrium} among all participants - such
that given that everyone else is telling the truth, no user has an incentive
to lie.

To see that our definition of incentive-compatibility is not vacuous, simply
add and subtract the term $X_{n+1}^{\prime}\hat{\beta}$ inside the squared
brackets on the left side term of (\ref{IC}), such that
\begin{align*}
E[R(X_{n+1})^{\prime}\hat{\beta}-X_{n+1}^{\prime}\beta_{0}]^{2} &
=E[R(X_{n+1})^{\prime}\hat{\beta}-X_{n+1}^{\prime}\hat{\beta}+X_{n+1}^{\prime
}\hat{\beta}-X_{n+1}^{\prime}\beta_{0}]^{2}\\
&  =E\left[  \Vert(R(X_{n+1})-X_{n+1})^{\prime}\hat{\beta}\Vert_{2}%
^{2}\right]  
 +E[X_{n+1}^{\prime}\hat{\beta}-X_{n+1}^{\prime}\beta_{0}]^{2}\\
&  +2E[\hat{\beta}^{\prime}(R(X_{n+1})-X_{n+1})X_{n+1}^{\prime}(\hat{\beta
}-\beta_{0})]\\
&  \geq E[X_{n+1}^{\prime}\hat{\beta}-X_{n+1}^{\prime}\beta_{0}]^{2}%
\end{align*}
Canceling common terms reduces incentive-compatibility to the following
inequality:
\begin{equation}
E\left[  \Vert(R(X_{n+1})-X_{n+1})^{\prime}\hat{\beta}\Vert_{2}^{2}\right]
\geq-2E[\hat{\beta}^{\prime}(R(X_{n+1})-X_{n+1})X_{n+1}^{\prime}(\hat{\beta
}-\beta_{0})].\label{modic}%
\end{equation}
Note that this inequality can go either way. For example, if all elements of
the vectors, $\hat{\beta},$ $X_{n+1}$ and $\hat{\beta}-\beta_{0}$ are
positive, and for every realized $R(X_{n+1})$, the difference $\tilde{X}%
_{n+1}-X_{n+1}$ is also positive, then incentive-compatibility holds. If,
however, $\hat{\beta}-\beta_{0}<0,$ while all the other terms are positive,
then incentive-compatibility may be violated. For instance, with one
explanatory variable, a very small lie can lead to a very small positive
number on the left hand side (due to a small lie being squared), while the
right hand side may be positive and slightly larger. For our main result, we
analyze the asymptotic version of incentive compatibility.

A weaker, ex-ante notion of incentive-compatibility considers a user, who
\textit{prior} to observing her covariates, commits to a strategy that maps
every possible realization of the covariates to a (possibly non-truthful)
report of these realized values. This notion fits situations in which the user
either automates her reports to the statistician, or delegates the reporting
to a third party. According to this notion, the estimator is ex-ante
incentive-compatible if on average (over the different realizations of the
user's covariates), the $n+1$ user has no incentive to misreport:
\[
\int E[(R(X_{n+1})^{\prime}\hat{\beta}-X_{n+1}^{\prime}\beta_{0}%
)^{2}]dP_{X_{n+1}}\geq\int E[(X_{n+1}^{\prime}\hat{\beta}-X_{n+1}^{\prime
}\beta_{0})^{2}]dP_{X_{n+1}},
\]
where the integral is computed with respect to the distribution of the new
user's attributes. Clearly, if an estimator is (ex-post) incentive-compatible,
then it is also ex-ante incentive compatible. Thus, the sufficient conditions
for (ex-post)\ incentive-compatibility of the Lasso estimator, which we
establish in Section 4, also guarantee ex-ante incentive-compatibility. While
ex-ante incentive-compatibility can be achieved with weaker conditions, the
proof of these conditions follows from our proof of (ex-post)
incentive-compatibility. In light of this, we shall focus on the ex-post
notion henceforth.

\subsection{Discussion}

In this subsection we discuss the motivation for some key ingredients of our
model, and we also remark on the implications of making alternative modeling
choices.\medskip

\noindent\textit{The choice of the Lasso estimator}. We chose to focus on
Lasso because it is the most basic machine learning technique that engages in
model selection. Since this is the first paper to ask, under what conditions
are such techniques incentive-compatible, it makes sense to start with the
most basic textbook technique. Once we understand whether and how to ensure
incentive compatibility in the simplest penalized regression model, we move on
to explore the weighted general penalized estimator (the Conservative Lasso)
in Section 5.

Nevertheless, it is worth mentioning that Lasso has several desirable
properties. First, its prediction error is of the same order of magnitude as
if there were an oracle, who could make predictions based on the true model.
This is shown in Theorem 6.4 and Corollary 6.3 of \cite{bvdg2011}, who provide
general oracle inequalities for convex loss with Lasso penalty. Second,
\cite{jwht2013} shows (see p.26) that despite being less flexible than
non-linear models such as random forests and deep learning, the Lasso
estimator can prevent overfitting, which is clearly a major issue in
out-of-sample contexts. In addition, Lasso is a continuous subset selection,
which has good prediction properties as shown in p.61-69 of \cite{htf2011}.

Since we also consider the Conservative Lasso in Section 5, we briefly mention
its properties here. Conservative Lasso is a two-step algorithm, where in the
first step, standard Lasso is run and all the variables are kept, and then in
the second step, a general weighted algorithm is run to select and estimate
the relevant variables. Conservative Lasso is therefore a general weighted
version of Lasso: when all weights are equal to one in the penalty, it reduces
to standard Lasso. Compared with the standard Lasso, the data-dependent
penalties of the Conservative Lasso allow for better differentiation of
relevant and irrelevant variables as seen in Lemma 1 of \cite{ck18}.
\footnote{The Adaptive Lasso is an alternative estimator that also uses a
data-dependent weighted penalty (see \cite{zou06}). However, in high
dimensional econometrics, the first step of the Adaptive Lasso can cut off
relevant variables, which can be undesirable as discussed in p.144-145 of
\cite{ck18}.} Further details on the Conservative Lasso will be provided in
Section 5.\medskip

\noindent\textit{The statistician's benevolence. }Our paper addresses the
issue raised in Eliaz and Spiegler (2019, 2020) that even if a statistician
wants to make the best prediction for the user (so there is no a priori
conflict of interest between them), the user may still have an incentive to
lie because of the model selection component in Lasso (or any penalized
regression for that matter), and because the user does not observe the
statistician's sample. Since the source of lying in this no-conflict benchmark
comes from the estimation procedure itself, the question is, how can we fix
the procedure - without harming its estimation properties - so as to ensure truth-telling?

What if the user and the statistician did have a conflict of interests - say,
the statistician uses the information that the user gives him in a way that
may harm the user? Then obviously, the user will have an incentive to lie no
matter which tuning parameter is chosen. In other words, in such an
environment, Lasso (or any other estimator) will not be incentive-compatible
unless the user is compensated, or the statistician uses an alternative
estimation technique that is not optimal econometrically (say, he deliberately
adds noise to it). Exploring this direction is clearly a separate research
agenda. \medskip

\noindent\textit{The user's loss function}. As explained above,
incentive-compatibility means that the user cannot profit by misreporting.
Suppose the user had a generic loss function $g(\cdot)$, such that
$Eg(R(X_{n+1}),X_{n+1})$ denoted the expected payoff of a user whose true
characteristics are given by $X_{n+1}$, but she reports the values
$R(X_{n+1})$. Then incentive-compatibility requires that $Eg(R(X_{n+1}%
),X_{n+1})\geq Eg(X_{n+1},X_{n+1})$ for any realization of $X_{n+1}$ and for
any report $R(X_{n+1})$. Note that in general, the user's expected payoff is
completely independent of the statistician's loss function. However, without
imposing any structure on $g(\cdot)$, it is impossible to characterize a
condition that ensures the incentive-compatibility of Lasso.

Given our focus on the no-conflict-of-interests benchmark (which we discussed
in the previous point), it is only natural to let the user and the
statistician have the same loss function that measures how far (in
expectation) the estimate is from the truth. For any loss function one chooses
for the statistician, the user has no incentive to lie if the expected loss
from lying (i.e., the distance between the predicted best outcome based on
lying and the actual ideal outcome for the agent) is higher than under
truth-telling. Hence, the definition of incentive-compatibility clearly
extends to any loss function shared by the statistician and the user. Of
course, for each candidate loss function one would need to find the exact
sufficient condition. We chose to focus on the mean squared error since it is the
most commonly used loss function.

If the user and the statistician evaluated the estimates using different loss
functions, then the incentive compatibility condition will apply only to the
user's loss function, and again, the precise sufficient condition for
incentive-compatibility will depend on the specification of this function.

\subsection{The Statistician's Data}


In this subsection, we introduce a number of restrictions on the
statistician's data. To describe these restrictions, we shall make use of the
following notations. Define an $l_{0}$ ball $\mathcal{B}_{l_{0}}%
(s_{0})=\{\Vert\beta_{0}\Vert_{l_{0}}\leq s_{0}\}$. Denote $\Sigma
:=EX_{i}X_{i}^{\prime}$ for $i=1,2,\cdots,n$, let $\hat{\Sigma}:=X^{\prime
}X/n$ be the sample counterpart, and let $\phi_{min}(\Sigma)$ denote the
minimum eigenvalue of $\Sigma.$ Our first requirement extends the sub-Gaussian
data assumption used in statistics:

\begin{assum}
\label{a1}(i).$E(u_{i}|X_{i})=0$,\quad$X_{i},u_{i}$\quad are\quad
identical\quad and\quad independent\quad across\quad$i=1,\cdots,n,$\quad
and\quad for\quad some\quad positive\quad constant\quad$C,$%
\begin{align*}
\max_{1\leq j\leq p}E|X_{ij}|^{4} &  \leq C<\infty\\
E|u_{i}|^{l} &  \leq C<\infty
\end{align*}
\linebreak where\quad$l=max(2k,4)$ for all $k\geq1$,\newline(ii). $\phi
_{min}(\Sigma)\geq c>0$, where $c$ is a positive constant.\bigskip
\end{assum}

Our second set of restrictions applies to the first and second moments. These
will guarantee the consistency of the Lasso estimator, but will not ensure
incentive compatibility (sufficient conditions for incentive compatibility
will be introduced in Section 4). We start by defining the maximal value of
certain cross products, which will be related to the behavior of moments in
high dimensions in our next assumption.%

\[
M_{1} := \max_{1 \le i \le n} \max_{1 \le j \le p} | X_{ij} u_{i}|,
\]

\[
M_{2} := \max_{1 \le i \le n} \max_{1 \le j \le p } \max_{1 \le l \le p} |
X_{il} X_{ij} - E X_{il} X_{ij}|.
\]

Note that $M_{1}$ is the maximal covariance between the regressors and errors
in a high dimensional context. Roughly speaking, when this covariance is
small, it captures exogeneity of the regressors in the sample. $M_{2}$ is the
maximal variance of the regressors in the sample. With large $p$ and $n$,
these covariance and variance terms can grow arbitrarily large - hence, we
need a condition that restricts the growth rate of their moments. Because we
are allowing for heteroskedastic data and unbounded regressors, we need to
consider the growth rate of \textit{higher-order} moments.
\footnote{Alternatively, we could provide primitives on Assumption \ref{a2}
using boundedness of individual moments of $X,u$.}{}

\begin{assum}
\label{a2}

(i).
\[
\frac{\sqrt{ln p}}{\sqrt{n}} [ max((E M_{1}^{2})^{1/2}, (E M_{2}^{2})^{1/2})]
\to0.
\]

(ii). $s_{0} (\frac{lnp}{n})^{1/2} \to0.$


(iii). $\| \beta_{0} \|_{2} = O (1)$.
\end{assum}

Assumption \ref{a2}(i) and \ref{a2}(ii) are standard in high dimensional
econometrics. In particular, \ref{a2}(i) is used in \cite{cck17} allowing them
to apply a concentration inequality, and \ref{a2}(ii) is a standard sparsity
condition. Note that with Assumption \ref{a2}(ii), Lasso prevents underfitting
since letting $\lambda=O(\sqrt{\frac{lnp}{n}})$ implies that $s_{0}\lambda
_{n}\rightarrow0,$ which ensures that $\lambda_{n}$ cannot be large enough to
generate underfitting. This allows us to establish the consistency of Lasso in
Lemmas A.1-A.3 in the Appendix.

Assumption \ref{a2}(iii) ensures that the signal to noise ratio is bounded
(see p.2343 of \cite{jvdg18}). To see this, set $\sigma_{u}^{2}:=var(u_{i})$,
the variance of the errors, such that $\sigma_{u}^{2}\geq c>0$, where $c$ is a
generic positive constant that is weakly below the minimum eigenvalue of
$\Sigma$ (which is positive by Assumption \ref{a1}). Hence, when
$E(u_{i}|X_{i})=0$, which is imposed in Assumption 1,
\[
\frac{var(y_{i})}{var(u_{i})}=\frac{\beta_{0}^{\prime}\Sigma\beta_{0}}%
{\sigma_{u}^{2}}+1,
\]
However,
\[
\frac{\beta_{0}^{\prime}\Sigma\beta_{0}}{\sigma_{u}^{2}}+1\geq\frac{\Vert
\beta_{0}\Vert_{2}^{2}\phi_{min}(\Sigma)}{\sigma_{u}^{2}}+1.
\]
where $\phi_{min}(\Sigma)\geq c>0$. Hence, if Assumption \ref{a2}(iii) holds,
then the signal to noise ratio satisfies $var(y_{i})/var(u_{i})\geq C_{0}%
+1>0$, with $C_{0}$ being a positive constant, and defined as $C_{0}%
:=\frac{\Vert\beta_{0}\Vert_{2}^{2}\phi_{min}(\Sigma)}{\sigma_{u}^{2}}$.

The empirical implication of this is that only a fixed number of nonzero
coefficients can be constants, and the other nonzero coefficients have to be
local to zero. To see this implication, note that
\[
\Vert\beta_{0}\Vert_{2}=\sqrt{\sum_{j=1}^{p}\beta_{0,j}^{2}}=\sqrt{\sum_{j\in
S_{0}}\beta_{0,j}^{2}}=O(1).
\]
since in the case of $s_{0}$ growing with $n$
\[
\sqrt{\sum_{j\in S_{0}}\beta_{0,j}^{2}}=\sqrt{\sum_{j\in F_{1}}\beta_{0,j}%
^{2}+\sum_{j\in S_{0}-F_{1}}\beta_{0,j}^{2}}\leq\sqrt{f_{1}C^{2}+(s_{0}%
-f_{1})\frac{C^{2}}{s_{0}-f_{1}}}=O(1),
\]
where $F_{1}:=\{j:|\beta_{0,j}|=C\}$ with $|F_{1}|=f_{1}$ being a fixed
number, $C$ is a generic positive constant and $F_{2}:=\{j:|\beta_{0,j}%
|=\frac{C}{\sqrt{s_{0}-f_{1}}}\}$ with $|F_{2}|=s_{0}-f_{1}$. For ease of
exposition, we set all coefficients in $F_{1}$ and $F_{2}$ to be the same
constants, $C$ and $C/\sqrt{s_{0}-f_{1}},$ respectively. $F_{2}$ contains
indices of all local to zero coefficients. This can easily be generalized
without affecting our results.

In Appendix B we take a more flexible approach compared with Assumption
\ref{a2}(iii). There, we assume that $\Vert\beta_{0}\Vert_{2}=O(\sqrt{s_{0}}%
)$. In this case, all nonzero coefficients can be large (i.e., none of them
are local to zero, as in set $F_{2}$ above). In other words, there is no index
set $F_{2}$ as above, but all nonzero coefficients (their indices) are in the
set $F_{1}$ above.




As $p$ and $n$ grow large, the total number of nonzero coefficients $s_{0}$
(also known as the \textit{sparsity index}) can grow arbitrarily large. To
guarantee consistency and unbiasedness, it is typically assumed that the
product of the sparsity index and the tuning parameter should go to zero.
However, this standard condition does not guarantee the incentive
compatibility of the Lasso estimator as can be seen in the proof of Theorem
\ref{t3} below.






\section{New Oracle Inequalities for Lasso}


Oracle inequalities in high dimensional statistics are upper bounds on
prediction and estimation errors. For our main result, we require moment
bounds on the Lasso estimator's error in $l_{1}$ norm. By taking the sample
size to be large, we can show that the upper bound on the mean of higher-order
moments of Lasso estimation errors tend to zero. We then use this asymptotic
result to establish the incentive compatibility of the Lasso estimator in
large samples. To illustrate this, we note that from the proof of Theorem
\ref{t3} in Appendix \ref{app4}, the incentive compatibility constraint is
tied to the following expression%
{\footnotesize
\begin{align}
E[R(X_{n+1})^{\prime}\hat{\beta}-X_{n+1}^{\prime}\beta_{0}]^{2}-E[X_{n+1}%
^{\prime}\hat{\beta}-X_{n+1}^{\prime}\beta_{0}]^{2} &  =E[\hat{\beta}^{\prime
}(R(X_{n+1})-X_{n+1})(R(X_{n+1})-X_{n+1})^{\prime}\hat{\beta}]\label{pt3.22}\\
&  +E[\hat{\beta}^{\prime}(R(X_{n+1})-X_{n+1})X_{n+1}^{\prime}(\hat{\beta
}-\beta_{0})]\label{pt3.23}\\
&  +E[(\hat{\beta}-\beta_{0})^{\prime}X_{n+1}(R(X_{n+1})^{\prime}%
-X_{n+1}^{\prime})\hat{\beta}].\label{pt3.24}%
\end{align}}
For incentive compatibility to hold in large samples, we need the sum of the
right-hand side terms to be greater than or equal to zero. The first term on
the right-hand side (\ref{pt3.22}) is always non-negative. Hence, if we prove
that (\ref{pt3.23}) and (\ref{pt3.24}) converge to zero, we establish
asymptotic incentive compatibility. However, the size of terms in
(\ref{pt3.23}) and (\ref{pt3.24}) will depend on the mean of higher-order
estimation errors of Lasso.

To bound these errors, we prove new oracle inequalities, which are different
from those that are given in the literature for $\Vert\hat{\beta}-\beta
_{0}\Vert_{1}$. These inequalities will serve an important role in proving our
main result in the next section (Theorem \ref{t3}). They are also of
independent interest as they extend previous results on sub-Gaussian data to
\textit{heteroskedastic} (conditionally) data sets that are commonly used in
econometrics. Our proof technique will also consider a less conservative bound
compared with \cite{jvdg18}. Hence, our new inequalities contribute to the
literature on high-dimensional econometrics where they can be used for proving
generalized semiparametric efficiency of Lasso-type-estimators (as, e.g., in
\cite{jvdg18}).

Our first result in this section is a $k$-th moment bound for the $l_{1}$ norm
of the Lasso bias. A key concept used in this result is the \textit{exception
probability} for the event $\mathcal{F}:=\{\mathcal{A}_{1}\cap\mathcal{A}%
_{2}\}$, where $\mathcal{A}_{1}$ and $\mathcal{A}_{2}$ are defined in
(\ref{ev1}) and (\ref{ev2}), which represent the empirical process-noise, and
the eigenvalue condition, respectively. The exception probability is the
complement of the event $\mathcal{F}$, and is denoted by $P(\mathcal{F}^{c})$.
An explicit upper bound for the exception probability is calculated in Lemma
\ref{l0}.

\begin{thm}
\label{t1}

Under Assumptions \ref{a1}-\ref{a2}, if $n$ is sufficiently large and
$\lambda_{n}\geq\frac{P(\mathcal{F}^{c})^{1/4k}}{s_{0}^{1/2}},$ then%

\[
[E \| \hat{\beta} - \beta_{0} \|_{1}^{k}]^{1/k} = O ( s_{0} \lambda_{n}).
\]



This result is valid uniformly over $\mathcal{B}_{l_{0}}(s_{0})=\{\Vert
\beta_{0}\Vert_{l_{0}}\leq s_{0}\}$.
\end{thm}

If we set $k=1$ we can learn whether the Lasso estimator is unbiased. By the
above Theorem, Assumption \ref{a2} and (\ref{ratel}) imply $s_{0}\lambda
_{n}\rightarrow0$. Hence, in large samples, we have unbiasedness in the large
$\lambda_{n}$ case. Next, we provide the $k$-th moment bound for $l_{1}$ norm
for the Lasso estimator.

\begin{thm}
\label{t2} Under Assumptions \ref{a1}-\ref{a2}, if $n$ is sufficiently large
$n$ and $\lambda_{n}\geq P(\mathcal{F}^{c})^{1/2k}/s_{0}^{1/2}$, then
\[
\lbrack E\Vert\hat{\beta}\Vert_{1}^{k}]^{1/k}=O(s_{0}^{1/2}).
\]


This result is valid uniformly over $\mathcal{B}_{l_{0}}(s_{0})=\{\Vert
\beta_{0}\Vert_{l_{0}}\leq s_{0}\}$.
\end{thm}

This is a new result and a simple extension of Theorem \ref{t1} above. The
rate in Theorem diverges to infinity if $s_{0}\rightarrow\infty$ as
$n\rightarrow\infty$.


\section{Incentive Compatibility of Lasso}

Our first main result, which is new in the literature on penalized
regressions, characterizes sufficient conditions for the Lasso estimator to be
incentive-compatible for a sufficiently large sample size. In other words, we
establish conditions such that when $n\rightarrow\infty,$
\[
E[R(X_{n+1})^{\prime}\hat{\beta}-X_{n+1}^{\prime}\beta_{0}]^{2}\geq
E[X_{n+1}^{\prime}\hat{\beta}-X_{n+1}^{\prime}\beta_{0}]^{2}.
\]
for all $X_{n+1}^{\prime}$ and $R(X_{n+1})^{\prime}$ and for every $\beta_{0}%
$, where the expectation is taken with respect to the statistician's realized
sample (since the reporting user does not observe this sample).

The proof of this result makes use of the following notation.%

\[
M_{3}:= \max_{1 \le j \le p} | X_{n+1,j}|,
\]

\[
M_{4}:\max_{1\leq j\leq p}|R(X_{n+1,j})-X_{n+1,j}|.
\]
Note that $M_{4}$ is the absolute magnitude of the potential misreport on a
given variable $j$ by the $n+1$ user. Since we deal with ex-post incentive
compatibility, $M_{3}$ and $M_{4}$ are deterministic but can grow with $n$.
Hence, we allow $M_{3}$ and $M_{4}$ to be nondecreasing in $n$.

\begin{thm}
\label{t3} Under Assumptions \ref{a1} and \ref{a2}, the Lasso estimator is
incentive compatible in large samples ($n\rightarrow\infty$) if the following
conditions hold:%
\begin{equation}
\lambda_{n}\geq P(\mathcal{F}^{c})^{1/8}/s_{0}^{1/2} \label{large lambda}%
\end{equation}
and
\begin{equation}
s_{0}^{3/2}\sqrt{\frac{lnp}{n}}[M_{3}][M_{4}]\rightarrow0. \label{C1}%
\end{equation}
Furthermore, incentive compatibility is valid uniformly over $\mathcal{B}%
_{l_{0}}(s_{0})=\{\Vert\beta_{0}\Vert_{l_{0}}\leq s_{0}\}$.
\end{thm}



\noindent\textbf{Remarks}.\medskip

\noindent\textbf{1.} Theorem \ref{t3} establishes that a \textit{sufficient}
condition for incentive compatibility is that the tuning parameter
$\lambda_{n}$ needs to be large \textquotedblleft enough\textquotedblright. A
simple way to choose $\lambda_{n}$ to satisfy (\ref{large lambda}) is to use
the upper bound of the exception probability
\[
\lambda_{n}:=upperbound(P(\mathcal{F}^{c})^{1/8}),
\]
in Lemma \ref{l0}. The simulations in Section 6 address the issue of whether
such a bound is feasible.\textbf{\medskip}

\noindent\textbf{2.} The typical concern with Lasso is the consistency of the
estimator ($\Vert\hat{\beta}-\beta_{0}\Vert_{1}=o_{p}(1)$), which can be
achieved by making sure that $\lambda_{n}$ goes to zero at a relatively fast
rate (as Lemma \ref{l1} in Appendix A shows, this rate is $s_{0}\lambda
_{n}\rightarrow0$). However, if $\lambda_{n}$ gets too small, the Lasso
estimator may admit many nonzero variables incorrectly (i.e., it creates an
\textit{overfit}). Consequently, when the number of regressors $p$ is very
large, the expectation of the sum of $l_{1}$ errors ($E\Vert\hat{\beta}%
-\beta_{0}\Vert_{1}$) can grow arbitrarily large, and incentive compatibility
may be violated. Put differently, \textit{consistency does not imply incentive
compatibility in large samples}. Thus, simply using the $l_{1}$ estimator
bound on its own does \textit{not} imply a bound for the \textit{expectation}
of $l_{1}$ error. 

We illustrate the point with a simple example. Suppose we take a value for
$\lambda_{n}$ below the upper bound in Theorem 1 (i.e., $\lambda
_{n}<P(\mathcal{F}^{c})^{1/4k}/s_{0}^{1/2}$). In particular, take $\lambda
_{n}=P(\mathcal{F}^{c})^{1/2k}/s_{0}^{1/2}$. Then from the proof of Theorem
1-(\ref{pt1.8}) we obtain that
\[
E\Vert\hat{\beta}-\beta_{0}\Vert_{1}^{k}=O(\lambda_{n}^{-k}P(\mathcal{F}%
^{c})^{1/2}),
\]
But given our choice of $\lambda_{n}$,
\[
\lambda_{n}^{-k}P(\mathcal{F}^{c})^{1/2}=[P(\mathcal{F}^{c})^{1/2k}%
/s_{0}^{1/2}]^{-k}P(\mathcal{F}^{c})^{1/2}=s_{0}^{k/2}\rightarrow\infty.
\]
Hence, even though there is consistency $(s_{0}\lambda_{n}\rightarrow0)$ under
this $\lambda_{n}$ choice (see Remark 5), the moment bound estimation error is
\textit{diverging}.
\[
E\Vert\hat{\beta}-\beta_{0}\Vert_{1}^{k}\rightarrow\infty.
\]

Why is overfitting a significant issue for incentive compatibility? The
intuition is as follows. Suppose the tuning parameter is sufficiently small so
that given the user's prior on the true coefficients, she expects that many
irrelevant variables will be included in the estimator. To correct this bias,
she can report that these variables are equal to zero.\textbf{\medskip}

\noindent\textbf{3.} The second sufficient condition (\ref{C1}) allows the
distance between the user's report $R (X_{n+1})$ and the truth $X_{n+1}$ to be
of any magnitude since $M_{4}\equiv\|R (X_{n+1})-X_{n+1}\Vert_{\infty}$ can be
arbitrarily large. Since the above conditions are sufficient but not
necessary, it remains an open question whether incentive compatibility can be
achieved with a tuning parameter that is lower than the threshold in
(\ref{large lambda}) without restricting the magnitude of the deviation
between the user's reported and true attributes.\textbf{\medskip}

\noindent\textbf{4.} Note that (\ref{C1}) requires stricter sparsity than
Assumption \ref{a2}. If $M_{3}=O(1)$ and $M_{4}=O(1)$, then condition
(\ref{C1}) amounts to $s_{0}^{3/2}\sqrt{\frac{ln\left(  p\right)  }{n}%
}\rightarrow0$, which is a sparsity requirement still stronger than Assumption
\ref{a2}(ii). In addition, if we let $M_{4}=O(ln\left(  n\right)  )$ and
$M_{3}=O(ln\left(  n\right)  ),$ then $s_{0}^{3/2}\sqrt{\frac{ln\left(
p\right)  }{n}}(ln\left(  n\right)  )^{2}\leq4s_{0}^{3/2}\sqrt{\frac{ln\left(
p\right)  }{n}}\rightarrow0$ is needed to get incentive compatibility with
$n\leq p$.\textbf{\medskip}

\noindent\textbf{5.} A natural question that arises is whether condition
(\ref{large lambda}) is compatible with the $l_{1}$ norm consistency of Lasso.
In other words, consistency requires a small $\lambda_{n}$, but incentive
compatibility requires a large $\lambda_{n}$, so are they compatible with each
other? When we select a large $\lambda_{n}$ to satisfy incentive
compatibility, we should not sacrifice consistency - i.e. we need
$s_{0}\lambda_{n}\rightarrow0$. To verify whether this is possible, we can
take the lower bound on the tuning parameter in (\ref{large lambda}) and see
whether we can achieve consistency. Note that
\begin{equation}
s_{0}\lambda_{n}=s_{0}\frac{P(\mathcal{F}^{c})^{1/8}}{s_{0}^{1/2}}=s_{0}%
^{1/2}P(\mathcal{F}^{c})^{1/8}, \label{r5.1}%
\end{equation}
From (\ref{pt1.8a}) in the Appendix, an upper bound on this exception
probability is:
\begin{equation}
P(\mathcal{F}^{c})\leq\frac{2}{p^{C_{1}}}+\frac{K[EM_{1}^{2}+EM_{2}^{2}%
]}{nlnp}, \label{r5.2}%
\end{equation}
where $C_{1}$ and $K$ are positive constants. With $l=1,2$, it therefore
follows from (\ref{r5.1}) and (\ref{r5.2}) that we need
\[
s_{0}^{4}/p^{C_{1}}\rightarrow0,\quad s_{0}^{4}\max_{l}EM_{l}^{2}%
/nlnp\rightarrow0,
\]
to have consistency. These two conditions are not unreasonable in the sense
that they are consistent with $\left(  n,p\right)  $ increasing to infinity.
Also they are compatible with moments satisfying condition (\ref{C1}) in
Theorem 3.\textbf{\medskip}

\noindent\textbf{6.} Finally, note that $\lambda_{n}=O(\sqrt{\frac{ln\left(
p\right)  }{n}})$ represents an upper bound in terms of rates for $\lambda
_{n}$, whereas (\ref{large lambda}) represents a lower bound. We can then take
for a positive constant $C>0$
\[
C\frac{\sqrt{ln(p)}}{\sqrt{n}}\geq\lambda_{n}\geq\frac{P(\mathcal{F}%
^{c})^{1/8}}{s_{0}^{1/2}}.
\]
The question is, are there suitable combinations of $n$ and $p$ that satisfy
these inequalities? By using algebra and the upper bound for exception
probability (\ref{pt1.8a}), we obtain the requirement that,
\[
Cs_{0}^{1/2}\geq\left[  \frac{2n}{p^{C_{1}}}+\frac{K[EM_{1}^{2}+EM_{2}^{2}%
]}{nlnp}\right]  ^{1/8}\frac{\sqrt{n}}{\sqrt{lnp}},
\]
which is plausible for $p>n$ and large $n$ since the left hand side may
diverge and the right side may go to zero. This may be the case for example
when $p$ is exponential in $n$, or large $C$.\textbf{\medskip}


\noindent\textbf{7.} When we relax Assumption 2(iii) to $\Vert\beta_{0}%
\Vert_{2}=O(\sqrt{s_{0}})$, the incentive compatibility is still satisfied but
under the slightly stronger condition
\[
s_{0}^{2}\sqrt{\frac{lnp}{n}}[M_{3}][M_{4}]\rightarrow0.
\]
The proofs are in Appendix B.2. Remarks 5-6 above still apply but with
slightly stronger sparsity conditions.
\textbf{\medskip}

\noindent\textbf{8}. While Theorem 3 provides sufficient conditions for
incentive-compatibility, we can also derive a necessary condition. As we show
in Appendix A, incentive-compatibility implies equation (A.52). By Markov's
inequality, we obtain that if the Lasso estimator is incentive-compatibile in
large samples, then
\[
\left[  \sum_{j=1}^{p}(\hat{\beta}_{j}-\beta_{0,j})X_{n+1,j}\right]  \left[
\sum_{j=1}^{p}(R(X_{n+1,j})-X_{n+1,j})\hat{\beta}_{j}\right]  \overset
{p}{\rightarrow}0,
\]
which implies a weighted consistency condition for the Lasso estimator. Note
that a condition like (\ref{large lambda}) is not involved, and hence, there
is still a gap between our sufficient and necessary conditions. It remains a
challenging open question whether there exist conditions for asymptotic
incentive compatibility of Lasso that are both necessary and sufficient.

\section{Incentive Compatibility Under a General Weighted Penalty: The
Conservative Lasso}

\label{iccl}

In this section we extend our analysis of incentive compatibility to a general
weighted penalty function. \cite{ck18} propose the Conservative Lasso, which
has superior model selection properties relative to the standard Lasso. This
is achieved by using a data-weighted penalty function. Specifically, the
Conservative Lasso is a two-step estimator
\[
\hat{\beta}_{w}=argmin_{\beta\in R^{p}}\{\Vert Y-X\beta\Vert_{n}^{2}%
+2\lambda_{n}\sum_{j=1}^{p}\hat{w}_{j}|\beta_{j}|\},
\]
where $Y=(y_{1},\cdots,y_{i},\cdots,y_{n})^{\prime}$ is an $n\times1$ vector,
$X$ is an $n\times p$ matrix, and with the prediction norm for a generic
vector $v$ defined as $\Vert v\Vert_{n}^{2}:=n^{-1}\sum_{i=1}^{n}v_{i}^{2}$.
The weights $\hat{w}_{j}$ are defined as follows: for each $j=1,\cdots,p$
\[
\hat{w}_{j}=\frac{\lambda_{prec}}{|\hat{\beta}_{j}|\cup\lambda_{prec}},
\]
where $\hat{\beta}_{j}$ is the Lasso estimator of Section 2 for variable $j$,
and $\lambda_{prec}$ is a positive sequence defined in Lemma \ref{lc1} in
Appendix C.

Roughly speaking, the Conservative Lasso may be viewed as giving excluded
variables in Lasso a \textquotedblleft second chance\textquotedblright. For
instance, when $\hat{\beta}_{j}=0$, the weight will be one (in contrast to a
weight of infinity in Adaptive Lasso). When $\hat{\beta}_{j}>\lambda_{prec}$,
the weight is less than one, so there is a differentiation of weights based on
Lasso estimation in the first step. A formal argument for weight properties,
and differentiation of relevant and irrelevant coefficients, is given in Lemma
1 of \cite{ck18}. For problems with the Adaptive Lasso in high dimensional
settings, see p.144-145 of \cite{ck18}.

In order to analyze the incentive-compatibility of the Conservative Lasso, we
will need the following assumption:

\begin{assum}
\label{as3}Define the precision matrix, $\Theta:=\Sigma^{-1}.$ Then

(i).
\[
\| \Theta\|_{l_{\infty}} = O(s_{1}),
\]
with $s_{1}$ a nondecreasing positive sequence in $n$.

(ii). $0< c \le\max_{1 \le j \le p} | \beta_{0,j}| \le C < \infty.$

(iii). $s_{1} \lambda_{n} = o(1)$.
\end{assum}

Assumption \ref{as3}(i) is a major relaxation of the assumptions in Lemma A.7
of \cite{ck18} (where it is used to derive $l_{\infty}$ bounds for
Conservative Lasso estimators) and in Lemma 4.1 of \cite{vdg16}. These papers
assume that the $l_{\infty}$ matrix norm of the precision matrix is constant,
which is quite restrictive since in many realistic environments, the dimension
of the matrix is $p\times p$ and its maximum row-sum can grow with $n$.

Assumption \ref{as3}(ii) prevents the maximum absolute coefficient from being
a sequence that is local to zero. A local to zero sequence for the maximum
coefficient is unrealistic, and furthermore, it implies that all the
coefficients in the model converge to zero, which renders the model useless to
begin with. Assumption \ref{as3}(iii) is needed for the minimum weights in the
Conservative Lasso to be bounded above by 1 (as prescribed by \cite{ck18}),
which constraints the growth rate of $s_{1}$ in Assumption \ref{as3}(i).

\begin{thm}
\label{t4} Under Assumptions \ref{a1}-\ref{as3}, with sufficiently large $n$,
and with
\[
\lambda_{n} \ge\frac{P (\mathcal{F}^{c})^{1/6k}}{s_{0}^{1/3} s_{1}^{1/3}}
\]
then we obtain
\[
\left[  E \| \hat{\beta}_{w} - \beta_{0} \|_{1}^{k}\right]  ^{1/k} = O (s_{0}
\lambda_{n}).
\]
The result is valid uniformly over $\mathcal{B}_{l_{0}} (s_{0})$.
\end{thm}

To the best of our knowledge, this is an entirely new result for general
weight functions such as the Conservative Lasso. This extends a result of
\cite{jvdg18} from Lasso with subgaussian data to Conservative Lasso with non
subgaussian data. The proofs are not trivial and involve finding the rate for
minimal estimated weight.

Note that the lower bound for the tuning parameter in Theorem \ref{t4} may be
weakly higher than the one in Theorem 1: if $s_{1}\leq\sqrt{s_{0}}$
\[
\frac{P(\mathcal{F}^{c})^{1/6k}}{s_{0}^{1/3}s_{1}^{1/3}}\geq\frac
{P(\mathcal{F}^{c})^{1/4k}}{s_{0}^{1/2}},
\]
since $s_{0}\geq1$. If, however, $s_{1}>\sqrt{s_{0}}$, then it is not clear
which bound will be higher.


Our next result, which is also new in the literature, provides a moment
estimator for the Conservative Lasso.

\begin{thm}
\label{t5} Under Assumptions \ref{a1}-\ref{as3}, with sufficiently large $n$,
and with
\[
\lambda_{n} \ge\frac{P (\mathcal{F}^{c})^{1/4k}}{s_{0}^{1/4} s_{1}^{1/2}}%
\]
then we obtain
\[
\left[  E \| \hat{\beta}_{w} \|_{1}^{k}\right]  ^{1/k} = O (s_{0}^{1/2}).
\]
The result is valid uniformly over $\mathcal{B}_{l_{0}} (s_{0})$.
\end{thm}

We are now ready to characterize the sufficient conditions for
incentive-compatibility in large samples of the Conservative Lasso.

\begin{thm}
\label{t6} Under Assumptions \ref{a1}-\ref{as3}, with sufficiently large $n$,
and with
\[
\lambda_{n} \ge max \left(  \frac{P (\mathcal{F}^{c})^{1/8}}{s_{0}^{1/4}
s_{1}^{1/2}}, \frac{P (\mathcal{F}^{c})^{1/12}}{s_{0}^{1/3} s_{1}^{1/3}%
}\right)  ,
\]
and
\[
s_{0}^{3/2} \frac{\sqrt{lnp}}{\sqrt{n}} M_{3} M_{4} \to0,
\]
then Conservative Lasso is Incentive Compatible in large samples. The result
is valid uniformly over $\mathcal{B}_{l_{0}} (s_{0})$.
\end{thm}

\textbf{Remarks.\medskip}

\noindent\textbf{1.} If our Assumption 3 imposed $s_{1}=O(1)$, as in
\cite{ck18}, then we would need a larger bound for the tuning parameter of the
Conservative Lasso compared with Lasso. This follows from observing that under
this restriction on $s_{1},$
\[
\frac{P(\mathcal{F}^{c})^{1/8}}{s_{0}^{1/2}}\leq\frac{P(\mathcal{F}^{c}%
)^{1/8}}{s_{0}^{1/4}s_{1}^{1/2}}.
\]
\textbf{\medskip}

\noindent\textbf{2.} A simple way to satisfy the lower bound for the tuning
parameter is to choose
\[
\lambda_{n}:={\mbox upperbound}\,(P(\mathcal{F}^{c})^{1/12}),
\]
by setting $s_{0}=1, s_{1}=1$.\textbf{\medskip}

\noindent\textbf{3. }A natural question that arises is whether a lower bound
for $\lambda_{n}$ is compatible with consistency as in Remark 5 of Theorem 3.
Applying the lower bound in Theorem 6, let
\[
s_{0}\lambda_{n}=s_{0}^{2/3}s_{1}^{-1/3}P(\mathcal{F}^{c})^{1/12}.
\]
Using (11) with $l=1,2$
\[
\frac{s_{0}^{8}}{p^{C_{1}}s_{1}^{4}}\rightarrow0,\quad\frac{s_{0}^{8}%
max_{l}EM_{l}^{2}}{nln\left(  p\right)  s_{1}^{4}}\rightarrow0.
\]
Note that the same exercise with the second bound in Theorem 6 results in
weaker conditions. These conditions are
\[
\frac{s_{0}^{6}}{p^{C_{1}}s_{1}^{4}}\rightarrow0,\,\frac{s_{0}^{6}\max
_{l}EM_{l}^{2}}{n(lnp)s_{1}^{4}}\rightarrow0.
\]
\textbf{\medskip}

\noindent\textbf{4. }Note that in a high dimensional penalized regression, the
tuning parameter $\lambda_{n}$ is an upper bound on the noise as defined by
$\mathcal{A}_{1}$ in (A.6). As in the case of Lasso (see Remark 6 following
Theorem 3), we verify whether the lower bound for incentive-compatibility is
compatible with the upper bound for noise reduction. Namely, we check whether
\[
C\frac{\sqrt{ln\left(  p\right)  }}{\sqrt{n}}\geq\lambda_{n}\geq
\frac{P(\mathcal{F}^{c})^{1/8}}{s_{0}^{1/4} s_{1}^{1/2}}.
\]
The question is, are there suitable combinations of $n$ and $p$ that satisfy
these inequalities? By using algebra and the upper bound for the exception
probability (\ref{pt1.8a}), we obtain the requirement that,
\[
Cs_{0}^{1/4} s_{1}^{1/2} \geq\left[  \frac{2n}{p^{C_{1}}}+\frac{K[EM_{1}%
^{2}+EM_{2}^{2}]}{nlnp}\right]  ^{1/8}\frac{\sqrt{n}}{\sqrt{lnp}},
\]
which is plausible for $p>n$ since the left-hand side may diverge and the
right-hand side may go to zero. For example, this may be the case when $p$ is
exponential in $n$. If, instead, we were to use
\[
C\frac{\sqrt{lnp}}{\sqrt{n}}\geq\lambda_{n}\geq\frac{P(\mathcal{F}^{c}%
)^{1/12}}{s_{0}^{1/3} s_{1}^{1/3}}.
\]
we would obtain the requirement that,
\[
Cs_{0}^{1/3} s_{1}^{1/3} \geq\left[  \frac{2n}{p^{C_{1}}}+\frac{K[EM_{1}%
^{2}+EM_{2}^{2}]}{nlnp}\right]  ^{1/12}\frac{\sqrt{n}}{\sqrt{lnp}},
\]
As before, this is plausible for $p>n$ since the left hand side may diverge
and the right side may go to zero (e.g., when $p$ is exponential in $n$). Note
that given that the upper bounds for $\lambda_{n}$ is the same, (see Lemma A.4
of \cite{ck18}), the lower bound for Conservative Lasso is weakly higher than
that of Lasso. Thus, the range of $(n,p)$ values that satisfy both bounds is
smaller in Conservative Lasso.
\textbf{\medskip}

\noindent\textbf{5.} We can also relax Assumption 2(iii) to $\Vert\beta
_{0}\Vert_{2}=O(\sqrt{s_{0}})$. The analysis will be similar to that of Lasso
(see Appendix B).

\section{Simulations}

This section has three objectives. First, it illustrates how in practice the
tuning parameter can be chosen to ensure incentive compatibility of the Lasso
estimator. Second, it demonstrates that by appropriately choosing the tuning
parameter (in line with the conditions in Theorem \ref{t3}), incentive
compatibility is analyzed through the lens of a small lie and larger lie.
Finally, we show that incentive compatibility is not vacuous, it is possible
to have new users lying to the machines and benefit from that.

We provide a simple simulation setup. Let
\[
y_{i}=X_{i}^{\prime}\beta_{0}+u_{i},
\]
where $\beta_{0}=(1,0_{p-s_{0}}^{\prime},1_{s_{0}-1}^{\prime})^{\prime}$,
$0_{p-s_{0}}$ is a $p-s_{0}$ column vector of all zero elements, and
$1_{s_{0}-1}$ is a $s_{0}-1$ dimensional column vector of all ones. Let
$s_{0}$ represent the sparsity of the above model and set $s_{0}=5$.

In our design we introduce a multivariate normal distribution for the
attributes of users $i=1,\cdots,n$, such that the covariance between the $j$
and $m$-th random variables are governed by
\[
\Sigma_{j,m}=0.5^{|j-m|},
\]
for $j=1,\dots,p$ and $m=1,\cdots,p$. Thus, the correlation between the
adjacent random variables is 0.5, and this declines when the random variables
are further apart. This Toeplitz type structure is commonly used in the high
dimensional literature (see \cite{ck18}). The new user has a draw from a $t$
distribution with three degrees of freedom. That new user's draw is
deterministic-non-random. It is drawn from $t_{3}$ and that is kept fixed
through the iterations so that we can compare between Lasso and Conservative
Lasso. The results are presented in Tables 1-4. Tables 1-2 consider Lasso with
a \textquotedblleft large\textquotedblright\ lie (the difference between the
truth and the new user's report is 2 across all attributes) and with a
\textquotedblleft small\textquotedblright\ lie (the difference between the
truth and thel report is 0.2 across all attributes). Tables 3-4 consider the
Conservative Lasso for the same setup.

For lasso, we aim to demonstrate that with a \textquotedblleft
large\textquotedblright\ tuning parameter as in Theorem \ref{t3}, incentive
compatibility can be achieved when the sample size $n$ is large enough. As
mentioned in the previous section, one possible choice of a tuning parameter
that satisfies Theorem \ref{t3} is the upper bound on the exception
probability,
\[
\lambda_{n}\geq upperbound(P(\mathcal{F}^{c})^{1/8}).
\]
The issue is to make the exception probability, $P(\mathcal{F}^{c})$
operational and usable. Note that an upper bound on this probability is (with
positive constants $C_{1}>0,C_{2}>0,K>0$)%

\begin{equation}
P(\mathcal{F}^{c})\leq\frac{2}{p^{C_{1}}}+\frac{K[EM_{1}^{2}+EM_{2}^{2}%
]}{nlnp}\leq\frac{2}{p^{C_{1}}}+\frac{C_{2}}{(lnp)^{2}}, \label{pfc1}%
\end{equation}
by observing that for $l=1,2$
\begin{align*}
\frac{K\max_{l}EM_{l}^{2}}{nlnp}  &  =\left[  \frac{K^{1/2}\sqrt{\max
_{l}EM_{l}^{2}}}{\sqrt{n}\sqrt{lnp}}\right]  ^{2}\\
&  =\left[  \frac{K^{1/2}\sqrt{\max_{l}EM_{l}^{2}}\sqrt{lnp}}{\sqrt{n}%
}\right]  ^{2}(\frac{1}{lnp})^{2}\\
&  \leq\frac{C_{2}}{(lnp)^{2}},
\end{align*}
where we use Assumption \ref{a2}(i). Hence, we can write the upper bound of
the exception probability by using $p\geq1$
\[
\frac{2}{p^{C_{1}}}+\frac{C_{2}}{(lnp)^{2}}\leq2+\frac{C_{2}}{(lnp)^{2}}.
\]
The tuning parameter is as follows%

\begin{equation}
\lambda_{n}:=\left[  2+\frac{C_{2}}{(lnp)^{2}}\right]  ^{1/8}, \label{lambdan}%
\end{equation}
where $C_{2}$ can start from a small positive value and stop at a large
positive value. We select the values for $C_{2}$ and $\lambda_{n}$ according
to the Generalized Information Criterion (GIC) as in Caner and Kock (2018),
which gives consistent model selection with weighted Lasso choices in the
least squares framework (the choice of tuning parameter with GIC in least
squares with Lasso and Conservative Lasso is shown to be consistent in Theorem
5 of Caner and Kock (2018)). Note that the criterion for choosing the tuning
parameter should take incentive compatibility into account. Hence, we choose
only $C_{2}$ with GIC, but the structure of our tuning parameter is determined
by our characterization of incentive compatibility. Therefore, our choice of
$\lambda_{n}$ is \textit{above} a lower bound, which prevents overfitting
(this is the novel insight of Theorem \ref{t3}). On the other hand, to prevent
a very large $\lambda_{n}$ and ensure consistency of Lasso, the lower bound
inversely depends on $p$.

Define%

\[
\lambda_{n}^{\ast}:=argmin_{\lambda_{n}\in\Lambda}\left[  ln(\hat{\sigma}%
^{2}(\lambda_{n}))+\frac{\hat{s}(\lambda_{n})}{n}ln(n)ln(ln(p))\right]  ,
\]
where $\hat{s}(\lambda_{n})$ is the number of nonzero elements in the Lasso
estimator, given a choice of $\lambda_{n}$ in a grid $\Lambda$, and
$\hat{\sigma}^{2}(\lambda_{n})$ is the mean squared residuals from the Lasso
regression, given a choice of $\lambda_{n}$ in a grid $\Lambda$. We form
$\Lambda$ as follows:\ we take $C_{2}$ in a grid of values $[2+\frac{C_{2}%
}{(lnp)^{2}}]$ as in (\ref{lambdan}). Let $C_{2}:=[0.01,0.1,0.5,1,2,10,100]$,
so $\Lambda$ is the grid of values of $\lambda_{n}$ depending on $C_{2}$. The
number of iterations is 1,000.

For the Conservative Lasso, the same type of tuning parameter analysis is
used, but with Theorem \ref{t6}, instead of Theorem 3. Hence, the tuning
parameter choice for conservative lasso is:%

\begin{equation}
\lambda_{n}:=\left[  2+\frac{C_{2}}{(lnp)^{2}}\right]  ^{1/12},
\label{lambdan-cl}%
\end{equation}
The Choice of $C_{2}$ is done in the same way as in lasso above. The
\textquotedblleft Report\textquotedblright\ column in Tables 1-4 display
$E[R(X_{n+1})^{\prime}\hat{\beta}-X_{n+1}^{\prime}\beta_{0}]^{2}$ as the mean
squared error from a false report by the user. \textquotedblleft
Truth\textquotedblright\ refers to $E[X_{n+1}^{\prime}(\hat{\beta}-\beta
_{0})]^{2}$. The difference between $R(X_{n+1})-X_{n+1}$ is kept at two
levels: 2 and 0.2 (for all $p$ variables), which represent large, and small
deviations from the truth. We have $p=100,200,300,$ and for each $p$ level we
analyze $n=100,200,300$.

The numbers in each cell of the tables correspond to the disutility of the
user (i.e., the mean square difference between the statistician's estimate and
the optimal action). Hence, smaller numbers correspond to higher payoffs. Let
us compare the tables when $p=300$ and $n=200.$ In Table 1, which corresponds
to a large magnitude of a lie, the user's disutility from reporting the truth
is 3.08, while the disutility from lying is 3.86. Hence, the $n+1$ user
prefers to be truthful. In Table 2, for a small lie, truth-telling induces a
disutility of 3.08, while lying induces a higher disutility of 2.04. Hence a
lie is preferred. Thus, even with our lower bound, it is possible to profit
from a \textquotedblleft small\textquotedblright\ lie. Note that some of the
small lies are prevented by our lower bound as can be seen in $p=200$ with
different $n$ in Table 1. So for small lies, guaranteeing
incentive-compatibility is more difficult. However, as predicted, all large
lies are prevented by our lower bound for the tuning parameter.

Tables 3-4 show the same pattern for Conservative Lasso. The lower bound on
the tuning parameter prevents large lies, but dissuading small lies depend on
$\left(  p,n\right)  $ combination. Also, when we move from small to large
lie, the mean squared error from lying gets very large. This is evident by
comparing Table 1 with Table 2, and comparing Table 3 with Table 4. To give an
example, for Conservative Lasso with $p=100\ $and $n=100$, in Table 3 the new
user prefers to lie with a mean squared error of 1.55 from lying compared to
2.49 from truth-telling. However, with a larger lie, the mean squared error
from lying increases to 5.45 making it not profitable to lie.

\begin{table}[ptb]%
\begin{tabular}
[c]{||c|c|c|c|c|c|c||}%
\multicolumn{7}{c}{Table 1: Lasso-Incentive Compatibility:}\\\hline\hline
Difference 2 & \multicolumn{2}{|c|}{$n=100$} & \multicolumn{2}{|c|}{$n=200$} &
\multicolumn{2}{|c|}{$n=300$}\\\hline
Dimension & Truth & Report & Truth & Report & Truth & Report\\\hline
$p=100$ & 2.71 & 4.25 & 2.85 & 3.54 & 2.72 & 3.71\\
$p=200$ & 0.99 & 18.01 & 0.76 & 17.95 & 0.68 & 17.93\\
$p=300$ & 3.35 & 4.22 & 3.08 & 3.86 & 3.05 & 3.61\\\hline
\end{tabular}
{\center
Note: "Truth" refers to $E [ X_{n+1}^{\prime}(\hat{\beta} - \beta_{0})]^{2} $
and "Report" refers to $E [ R (X_{n+1})^{\prime}\hat{\beta} - X_{n+1}^{\prime
}\beta_{0}]^{2} $ \linebreak in Incentive Compatibility Definition. Smaller
values of these average squared errors are desirable.}\end{table}

\begin{table}[ptb]%
\begin{tabular}
[c]{||c|c|c|c|c|c|c||}%
\multicolumn{7}{c}{Table 2:Lasso-Incentive Compatibility:}\\\hline\hline
Difference 0.2 & \multicolumn{2}{|c|}{$n=100$} & \multicolumn{2}{|c|}{$n=200$}
& \multicolumn{2}{|c|}{$n=300$}\\\hline
Dimension & Truth & Report & Truth & Report & Truth & Report\\\hline
$p=100$ & 2.71 & 1.80 & 2.85 & 1.90 & 2.71 & 1.76\\
$p=200$ & 0.99 & 1.64 & 0.76 & 1.40 & 0.68 & 1.32\\
$p=300$ & 3.35 & 2.32 & 3.08 & 2.04 & 3.05 & 2.00\\\hline
\end{tabular}
{\center
Note: "Truth" refers to $E [ X_{n+1}^{\prime}(\hat{\beta} - \beta_{0})]^{2} $
and "Report" refers to $E [ R (X_{n+1})^{\prime}\hat{\beta} - X_{n+1}^{\prime
}\beta_{0}]^{2} $ \linebreak in Incentive Compatibility Definition. Smaller
values of these average squared errors are desirable.}\end{table}

\newpage

\begin{table}[ptb]%
\begin{tabular}
[c]{||c|c|c|c|c|c|c||}%
\multicolumn{7}{c}{Table 3: Conservative Lasso-Incentive Compatibility:}%
\\\hline\hline
Difference 2 & \multicolumn{2}{|c|}{$n=100$} & \multicolumn{2}{|c|}{$n=200$} &
\multicolumn{2}{|c|}{$n=300$}\\\hline
Dimension & Truth & Report & Truth & Report & Truth & Report\\\hline
$p=100$ & 2.49 & 5.45 & 2.56 & 4.78 & 2.43 & 4.85\\
$p=200$ & 0.89 & 19.78 & 0.70 & 19.48 & 0.63 & 19.47\\
$p=300$ & 3.06 & 5.27 & 2.79 & 4.94 & 2.78 & 4.65\\\hline
\end{tabular}
{\center
Note: "Truth" refers to $E [ X_{n+1}^{\prime}(\hat{\beta} - \beta_{0})]^{2} $
and "Report" refers to $E [ R (X_{n+1})^{\prime}\hat{\beta} - X_{n+1}^{\prime
}\beta_{0}]^{2} $ \linebreak in Incentive Compatibility Definition. Smaller
values of these average squared errors are desirable.}\end{table}

\begin{table}[ptb]%
\begin{tabular}
[c]{||c|c|c|c|c|c|c||}%
\multicolumn{7}{c}{Table 4: Conservative Lasso-Incentive Compatibility:}%
\\\hline\hline
Difference 0.2 & \multicolumn{2}{|c|}{$n=100$} & \multicolumn{2}{|c|}{$n=200$}
& \multicolumn{2}{|c|}{$n=300$}\\\hline
Dimension & Truth & Report & Truth & Report & Truth & Report\\\hline
$p=100$ & 2.49 & 1.55 & 2.56 & 1.58 & 2.43 & 1.47\\
$p=200$ & 0.89 & 1.57 & 0.70 & 1.36 & 0.63 & 1.29\\
$p=300$ & 3.06 & 2.03 & 2.79 & 1.75 & 2.78 & 1.72\\\hline
\end{tabular}
{\center
Note: "Truth" refers to $E [ X_{n+1}^{\prime}(\hat{\beta} - \beta_{0})]^{2} $
and "Report" refers to $E [ R (X_{n+1})^{\prime}\hat{\beta} - X_{n+1}^{\prime
}\beta_{0}]^{2} $ \linebreak in Incentive Compatibility Definition. Smaller
values of these average squared errors are desirable.}\end{table}

\newpage

\section{Conclusion}

The growing reliance on machine learning in automating decisions previously
made by people raises the question of how people would interact with these
automated systems. In particular, would people have an incentive to act
strategically in order to manipulate such automated systems? This strategic
interaction will become particularly important when these automated systems
start playing a more prominent role in medical decision-making or even in driving.

This paper takes only a small preliminary step towards addressing this
question by studying whether a user would want to lie to an automated system
that uses Lasso or Conservative Lasso to predict that user's ideal outcome
based on her reported attributes. Our main contribution is showing that
truthful reporting can be ensured by appropriately adjusting the tuning
parameter to be larger than what is required for consistency. Our result is
also significant from a pure econometrics point of view: just concentrating on
oracle inequalities and post-selection inference can lead to a small tuning
parameter, which in turn, can lead to model overfitting, which then introduces
an incentive to misreport. If users have an incentive to provide false input
to algorithms used for estimation and prediction, then it is no longer clear
that one can rely on the output of these algorithms.






\setcounter{equation}{0}\setcounter{lemma}{0}\renewcommand{\theequation}{A.\arabic{equation}}\renewcommand{\thelemma}{A.\arabic{lemma}}

\appendix

In the next part, Appendix A considers the proofs when $p>n$, and Appendix B
considers the case $p\le n$, and relaxing Assumption \ref{a2}(iii). Appendix C
covers Conservative Lasso proofs.

\section{Appendix A}

\subsection{Notation}

In this section, we show some results that will help us in proofs. Define
random vector of variables $F_{i} := ( F_{i1}, \cdots, F_{ij}, \cdots,
F_{ip})^{\prime}$. Also define $\sigma_{F}^{2} := n (max_{1 \le j \le p} var
F_{ij})$, and $M_{F} := \max_{1 \le i \le n} \max_{1 \le j \le p} |F_{ij} - E
F_{ij} |$. Note that $\hat{\mu}_{j} := n^{-1} \sum_{i=1}^{n} F_{ij}$, and
$\mu_{j} := E F_{ij}$.

\subsection{Maximal Inequalities}

We use two assumptions that will provide us maximal inequalities.

\textbf{Assumption A.1}. Assume $F_{i}$ are iid random vectors across
$i=1,2,\cdots, n$ with $\max_{1 \le j \le p} var F_{ij} \le C < \infty$ for a
positive constant $C >0$.\newline

\textbf{Assumption A.2}. Assume
\[
\frac{\sqrt{E M_{F}^{2}} \sqrt{ln p}}{\sqrt{n}} \to0.
\]

We use the following maximal inequality. With Assumption A.1, Lemma E.2(ii) of
\cite{cck17} is: (see (A.2) of \cite{ck19})
\begin{equation}
P \left[  \max_{1 \le j \le p} | \hat{\mu}_{j} - \mu_{j} | \ge2 E \max_{1 \le
j \le p} |\hat{\mu}_{j} - \mu_{j} | + \frac{t}{n} \right]  \le exp(-t^{2}/3
\sigma_{F}^{2}) + K \frac{ E M_{F}^{2}}{t^{2}}, \label{sa1}%
\end{equation}
for a constant $K >0$. With Assumptions A.1-A.2 here, \cite{ck19} or Lemma E.1
of \cite{cck17} provides
\begin{align}
E \max_{1 \le j \le p} | \hat{\mu}_{j} - \mu_{j} |  &  \le K [ \frac
{\sqrt{lnp}}{\sqrt{n}} + \frac{ \sqrt{E M_{F}^{2}} ln p}{n}]\nonumber\\
&  = O ( \frac{\sqrt{ln p}}{\sqrt{n}}). \label{sa2}%
\end{align}

\noindent Define the sequence $\kappa_{n} = ln p$. Set $t = t_{n} = ( n
\kappa_{n})^{1/2}$ to have (\ref{sa1}) as
\begin{align}
P \left[  \max_{1 \le j \le p} | \hat{\mu}_{j} - \mu_{j} | \ge2 E \max_{1 \le
j \le p} | \hat{\mu}_{j} - \mu_{j} | + \frac{\sqrt{\kappa_{n}}}{\sqrt{n}}
\right]   &  \le exp(- C_{1} \kappa_{n} ) + K \frac{ E M_{F}^{2}}{n \kappa
_{n}}\nonumber\\
&  = \frac{1}{p^{C_{1}}} + \frac{K E M_{F}^{2}}{n ln p} \label{sa3}%
\end{align}
where $C_{1}>0$, is a positive constant.

Now combine (\ref{sa2}) with (\ref{sa3}) to have%

\begin{align}
P ( \max_{1 \le j \le p} | \hat{\mu}_{j} - \mu_{j} |  &  \ge2K [ \frac
{\sqrt{lnp}}{\sqrt{n}} + \frac{ (E M_{F}^{2})^{1/2} ln p}{n}] + \frac
{\sqrt{lnp}}{\sqrt{n}} )\nonumber\\
&  \le\frac{1}{p^{C_{1}}} + \frac{ K E M_{F}^{2}}{n (ln p)} = o(1),
\label{sa4}%
\end{align}
by Assumptions A1-A.2. This shows also that, since $E M_{F}^{2}$ is
nondecreasing in $n$
\begin{equation}
\max_{1 \le j \le p } | \hat{\mu}_{j} - \mu_{j} | = O_{p} ( \sqrt{lnp}%
/\sqrt{n}). \label{sa5}%
\end{equation}

\subsubsection{Events}

Before the assumptions, we need to define events that will be helpful. The
first event is:
\begin{equation}
\mathcal{A}_{1} =\left\{  2 \left\|  \frac{u^{\prime}X}{n} \right\|  _{\infty}
\le\lambda_{n} \right\}  , \label{ev1}%
\end{equation}
which controls the noise. This is the maximal correlation between regressors
and errors. We want this to be bounded with probability approaching one, and
this upper bound, $\lambda_{n}$, itself is converging to zero in our proofs.
We show that in Lemma A.2. So in large samples, this proof technique amounts
to verification of exogeneity of regressors. This is standard in high
dimensional econometrics, for a recent analysis see Lemma A.4 of \cite{ck18}.

We start with defining first population counterparts of restricted eigenvalue
conditions and then show the empirical version also. These are standard in
high dimensional econometrics and statistics and can be seen from Assumption 1
of \cite{ck18}.

We define the population adaptive restricted eigenvalue of $\Sigma$
\begin{equation}
\phi_{\Sigma}^{2} (s) = \min\left\{  \frac{\delta^{\prime}\Sigma\delta}{\|
\delta_{S} \|_{2}^{2}}: \delta\in R^{p} - \{0\}, \| \delta_{S_{c}} \|_{1} \le3
\sqrt{s} \| \delta_{S} \|_{2}, | S | \le s \right\}  . \label{pope1}%
\end{equation}

\noindent Note that if $\Sigma= E X_{i} X_{i}^{\prime}$ has full rank, the
population adaptive restricted eigenvalue being positive is satisfied by
Assumption \ref{a1}. Also instead of minimizing all over $R^{p}$, we minimize
vectors that satisfy $\| \delta_{S^{c}} \|_{1} \le3 \| \delta_{S} \|_{1}$.
Even in the cases that $\Sigma$ does not have full rank, it is possible that
minimal adaptive restricted eigenvalue condition is satisfied due to
optimization over a restricted set. The parameter $\delta$ will be related to
structural parameter $\beta$ in the proofs.

First define the empirical adaptive restricted eigenvalue condition, which is
empirical counterpart of the population version in Assumption \ref{a1}:
\begin{equation}
\hat{\phi}_{\hat{\Sigma}}^{2} (s) = \min\left\{  \frac{\delta^{\prime}%
\hat{\Sigma} \delta}{\| \delta_{S} \|_{2}^{2}}: \delta\in R^{p} - \{0\}, \|
\delta_{S_{c}} \|_{1} \le3 \sqrt{s} \| \delta_{S} \|_{2}, | S | \le s
\right\}  . \label{eev}%
\end{equation}
We are interested in behavior of the minimal empirical adaptive restricted
eigenvalue condition evaluated for set $S_{0}$ at cardinality $s_{0}$. The
second event is:
\begin{equation}
\mathcal{A}_{2} = \left\{  \hat{\phi}_{ \hat{\Sigma} }^{2} (s_{0}) \ge
\phi_{\Sigma}^{2} (s_{0})/2 \right\}  . \label{ev2}%
\end{equation}
Empirical adaptive restricted eigenvalue condition is needed since in case of
$p>n$, $X^{\prime}X$ is singular and the minimal eigenvalue of $X^{\prime}X$
is zero. Empirical adaptive eigenvalue is over a restricted set which we prove
to be positive, with probability approaching one, in Lemma A.3. This is also
standard in high dimensional econometrics, see Lemma A.6 of \cite{ck18}. Set
$\mathcal{F} = \mathcal{A}_{1} \cap\mathcal{A}_{2}$, and the complement event
as $\mathcal{F}^{c}$.

\subsubsection{Proofs of Lemmata}

The following four Lemmata are the intermediate results that are used for Theorems.

\begin{lemma}
\label{l1} Under the joint event $\mathcal{F}:=\{ \mathcal{A}_{1}
\cap\mathcal{A}_{2} \}$ we have
\[
\| \hat{\beta} - \beta_{0} \|_{1} \le\frac{24 \lambda_{n} s_{0} }%
{\phi_{{\Sigma}}^{2} (s_{0})}.
\]
This is also valid uniformly over $\mathcal{B}_{l_{0}} (s_{0}) = \{ \|
\beta_{0} \|_{l_{0}} \le s_{0} \}$.
\end{lemma}

\noindent\textbf{Proof of Lemma \ref{l1}}. Using $\hat{\beta}$ definition%

\[
\| Y - X \hat{\beta} \|_{n}^{2} + 2 \lambda_{n} \sum_{j=1}^{p} | \hat{\beta
}_{j}| \le\| Y - X \beta_{0} \|_{n}^{2} + 2 \lambda_{n} \sum_{j=1}^{p} |
\beta_{0,j}|.
\]
Use the model $Y= X \beta_{0} + u$ on the first left side term as well as the
first right side term to simplify the inequality above combining with Holder's
Inequality
\begin{align*}
\| X (\hat{\beta} - \beta_{0} ) \|_{n}^{2} + 2 \lambda_{n} \sum_{j=1}^{p} |
\hat{\beta}_{j} |  &  \le2 \left|  \frac{u^{\prime}X}{n} (\hat{\beta} -
\beta_{0}) \right|  + 2 \lambda_{n} \sum_{j=1}^{p} | \beta_{0,j}|\\
&  \le2 \| \frac{u^{\prime}X}{n} \|_{\infty} \| \hat{\beta} - \beta_{0} \|_{1}
+ 2 \lambda_{n} \sum_{j=1}^{p} | \beta_{0,j}|
\end{align*}
On the right side assuming we are on the event $\mathcal{A}_{1}$
\[
2 \| \frac{u^{\prime}X}{n} \|_{\infty} \| \hat{\beta} - \beta_{0} \|_{1}
\le\lambda_{n} \| \hat{\beta} - \beta_{0} \|_{1}.
\]
So we have
\[
\| X (\hat{\beta} - \beta_{0} ) \|_{n}^{2} + 2 \lambda_{n} \sum_{j=1}^{p} |
\hat{\beta}_{j} | \le\lambda_{n} \| \hat{\beta} - \beta_{0} \|_{1} + 2
\lambda_{n} \sum_{j=1}^{p} | \beta_{0,j}|.
\]
Use $\| \hat{\beta} \|_{1} = \| \hat{\beta}_{S_{0}} \|_{1} + \| \hat{\beta
}_{{S_{0}}^{c}} \|_{1}$ on the second term for the left side of the inequality
immediately above
\[
\| X (\hat{\beta} - \beta_{0} ) \|_{n}^{2} + 2 \lambda_{n} \sum_{j \in
S_{0}^{c}} | \hat{\beta}_{j} | \le\lambda_{n} \| \hat{\beta} - \beta_{0}
\|_{1} + 2 \lambda_{n} \sum_{j=1}^{p} | \beta_{0,j}| - 2 \lambda_{n} \sum_{j
\in S_{0}} | \hat{\beta}_{j} |.
\]
By assumption of sparsity $\sum_{j \in S_{0}^{c}} | \beta_{0,j}| =0$, and
using the reverse triangle inequality we have
\[
\| X (\hat{\beta} - \beta_{0} ) \|_{n}^{2} + 2 \lambda_{n} \sum_{j \in
S_{0}^{c}} | \hat{\beta}_{j} | \le\lambda_{n} \| \hat{\beta} - \beta_{0}
\|_{1} + 2 \lambda_{n} \sum_{j \in S_{0}} | \hat{\beta}_{j} - \beta_{0,j} |.
\]

Next by $\| \hat{\beta} - \beta_{0} \|_{1} = \| \hat{\beta}_{S_{0}} -
\beta_{0,S_{0}} \|_{1} + \| \hat{\beta}_{S_{0}^{c}} \|_{1}$ for the first term
on the right side of the inequality immediately above
\[
\| X (\hat{\beta} - \beta_{0} ) \|_{n}^{2} + \lambda_{n} \sum_{j \in S_{0}%
^{c}} | \hat{\beta}_{j} | \le3 \lambda_{n} \sum_{j \in S_{0}} | \hat{\beta
}_{j} - \beta_{0,j}|.
\]
Use $\| \hat{\beta}_{S_{0}} - \beta_{0, S_{0}} \|_{1} \le\sqrt{s_{0}} \|
\hat{\beta} - \beta_{0, S_{0}} \|_{2}$ above on the right side to have
\begin{equation}
\| X (\hat{\beta} - \beta_{0} ) \|_{n}^{2} + \lambda_{n} \sum_{j \in S_{0}%
^{c}} | \hat{\beta}_{j} | \le3 \lambda_{n} \sqrt{s_{0}} \| \hat{\beta}_{S_{0}}
- \beta_{0, S_{0}} \|_{2}. \label{1}%
\end{equation}

\noindent Ignoring the first term on the left of (\ref{1}), (\ref{1}) shows
that we satisfy the restricted set condition in empirical adaptive restricted
eigenvalue condition, so we have
\[
\| \hat{\beta}_{{S_{0}}^{c} } \|_{1} \le3 \sqrt{s_{0}} \| \hat{\beta}_{S_{0}}
- \beta_{0, S_{0}} \|_{2}.
\]
Using $\delta= \hat{\beta} - \beta_{0}$ in the empirical adaptive restricted
eigenvalue condition (\ref{eev}) in (\ref{1})
\[
\| X (\hat{\beta} - \beta_{0} ) \|_{n}^{2} + \lambda_{n} \sum_{j \in S_{0}%
^{c}} | \hat{\beta}_{j} | \le3 \lambda_{n} \sqrt{s_{0}} \frac{ \| X^{\prime
}(\hat{\beta} - \beta_{0})\|_{n}}{\hat{\phi}_{\hat{\Sigma}} (s_{0})}.
\]
Then use $3 uv \le u^{2}/2+ 9v^{2}/2$ with $u= \lambda_{n} \sqrt{s_{0}} /
\hat{\phi}_{\hat{\Sigma}} (s_{0})$, $v = \| X (\hat{\beta} - \beta_{0})\|_{n}$
to get
\[
\| X (\hat{\beta} - \beta_{0} ) \|_{n}^{2} + \lambda_{n} \sum_{j \in S_{0}%
^{c}} | \hat{\beta}_{j} | \le\frac{ \| X (\hat{\beta} - \beta_{0}) \|_{n}^{2}%
}{2} + \frac{9}{2} \frac{\lambda_{n}^{2} s_{0}}{\hat{\phi}_{ \hat{\Sigma}}^{2}
(s_{0})}.
\]
Simplify above
\[
\| X (\hat{\beta} - \beta_{0} ) \|_{n}^{2} + 2 \lambda_{n} \sum_{j \in
S_{0}^{c}} | \hat{\beta}_{j} | \le\frac{ 9 \lambda_{n}^{2} s_{0}}{\hat{\phi}_{
\hat{\Sigma}}^{2} (s_{0})}.
\]
Use the event $\mathcal{A}_{2}$ we get the following
\[
\| X (\hat{\beta} - \beta_{0} ) \|_{n}^{2} + 2 \lambda_{n} \sum_{j \in
S_{0}^{c}} | \hat{\beta}_{j} | \le\frac{ 18 \lambda_{n}^{2} s_{0}}{\phi_{
\Sigma}^{2} (s_{0})}.
\]
This implies the oracle inequality
\begin{equation}
\| X (\hat{\beta} - \beta_{0} ) \|_{n}^{2} \le\frac{ 18 \lambda_{n}^{2} s_{0}%
}{\phi_{ \Sigma}^{2} (s_{0})}. \label{2}%
\end{equation}

To get to the $l_{1}$ bound ignore the first term in (\ref{1}) and add both
sides $\lambda_{n} \| \hat{\beta}_{S_{0}} - \beta_{0, S_{0}} \|_{1}$ to have
\[
\lambda_{n} \sum_{j \in S_{0}^{c}} | \hat{\beta}_{j} | + \lambda_{n} \sum_{j
\in S_{0}} | \hat{\beta}_{j} - \beta_{0,j}|= \lambda_{n} \| \hat{\beta} -
\beta_{0} \|_{1} \le\lambda_{n} \| \hat{\beta}_{S_{0}} - \beta_{0, S_{0}}
\|_{1} + 3 \lambda_{n} \sqrt{s_{0}} \| \hat{\beta}_{S_{0}} - \beta_{0, S_{0}}
\|_{2},
\]
by seeing also $\sum_{j \in S_{0}^{c}} |\beta_{0,j}| = 0$. Now use the norm
inequality $\| \hat{\beta}_{S_{0}} - \beta_{0, S_{0}} \|_{1} \le\sqrt{s_{0}}
\| \hat{\beta}_{S_{0}} - \beta_{0, S_{0}} \|_{2}$ to have
\[
\lambda_{n} \| \hat{\beta} - \beta_{0} \|_{1} \le4 \lambda_{n} \sqrt{s_{0}} \|
\hat{\beta}_{S_{0}} - \beta_{0, S_{0}} \|_{2}.
\]
Use the empirical adaptive restricted eigenvalue condition with $\delta=
\hat{\beta} - \beta_{0}$
\[
\| \hat{\beta} - \beta_{0} \|_{1} \le4 \sqrt{s_{0}} \frac{ \| X (\hat{\beta} -
\beta_{0}) \|_{n}}{\hat{\phi}_{\hat{\Sigma}} (s_{0})}.
\]
Use (\ref{2}) and the event $\mathcal{A}_{2}$ to have
\begin{align}
\| \hat{\beta} - \beta_{0} \|_{1}  &  \le4 \sqrt{s_{0}} \left[  \frac{3
\sqrt{2} \lambda_{n} \sqrt{s_{0}}}{\phi_{\Sigma} (s_{0})} \right]  \left[
\frac{1}{\hat{\phi} _{\hat{\Sigma}} (s_{0})}\right] \nonumber\\
&  \le\frac{24 \lambda_{n} s_{0}}{\phi_{\Sigma}^{2} (s_{0})}. \label{3}%
\end{align}
Note that uniformity over $\mathcal{B}_{l_{0}} (s_{0})$ follows since the
upper bound in (\ref{3}) depends on $\beta_{0}$ only through $s_{0}$.
\textbf{Q.E.D}

\begin{lemma}
\label{l2} (i). Under Assumption \ref{a1}, and since $\kappa_{n} = ln p$
\[
P (\mathcal{A}_{1} ) \ge1 - exp (-C_{1} \kappa_{n}) - \frac{K E M_{1}^{2}}{(n
\kappa_{n})} = 1 - \frac{1}{p^{C_{1}}} - \frac{K E M_{1}^{2}}{n ln p}
\]

(ii). Under added Assumption \ref{a2} to Assumption \ref{a1}, $P (
\mathcal{A}_{1}) \to1$.

(iii). Under added Assumption \ref{a2} to Assumption \ref{a1}, $\lambda_{n} =
O (\sqrt{lnp/n})$.
\end{lemma}

\noindent\textbf{Proof of Lemma \ref{l2}}. (i). Establish the probability
bound on $\mathcal{A}_{1}$ via Assumption \ref{a1}, using (\ref{sa3}%
)(\ref{sa4}) with $F_{i}= X_{i} u_{i}$ there and $\kappa_{n} = ln p$, we have
\begin{equation}
P (\mathcal{A}_{1} ) \ge1 - exp (-C_{1} \kappa_{n}) - K \frac{E M_{1}^{2}}{(n
\kappa_{n})} =1 - \frac{1}{p^{C_{1}}} - \frac{K E M_{1}^{2}}{n ln p},
\label{4}%
\end{equation}
with
\begin{equation}
\lambda_{n} = K [ \sqrt{\frac{lnp}{n}} + \frac{ \sqrt{E M_{1}^{2}} ln p}{n}] +
\sqrt{\frac{ln p}{n}}. \label{lambda}%
\end{equation}

(ii). By Assumption \ref{a2}, we have the proof.

(iii). By Assumption \ref{a2}, we have
\begin{equation}
\lambda_{n} = O ( \sqrt{lnp/n}). \label{ratel}%
\end{equation}
\textbf{Q.E.D.}

\begin{lemma}
\label{l3} Under Assumptions \ref{a1}, \ref{a2}, $\kappa_{n} = ln p$
\[
P (\mathcal{A}_{2}) \ge1 - exp (- C_{1} \kappa_{n}) - \frac{K E M_{2}^{2}}{(n
\kappa_{n})} = 1 - \frac{1}{p^{C_{1}}} - \frac{ K E M_{2}^{2}}{n ln p} = 1-
o(1).
\]

\end{lemma}

\noindent\textbf{Proof of Lemma \ref{l3}}. Start with
\begin{align}
\left|  \delta^{\prime}\frac{X^{\prime}X}{n} \delta\right|   &  = \left|
\delta^{\prime}( \frac{X^{\prime}X}{n} - \Sigma+ \Sigma) \delta\right|
\nonumber\\
&  \ge| \delta^{\prime}\Sigma\delta| - | \delta^{\prime}(\hat{\Sigma} -
\Sigma) \delta|. \label{e1}%
\end{align}
The second term on the right side of (\ref{e1}) can be bounded by repeated
application of Holders inequality
\[
| \delta^{\prime}(\hat{\Sigma} - \Sigma) \delta| \le\| \delta\|_{1}^{2} \|
\hat{\Sigma} - \Sigma\|_{\infty}.
\]
So (\ref{e1}) becomes
\begin{equation}
| \delta^{\prime}\hat{\Sigma} \delta| \ge| \delta^{\prime}\Sigma\delta| - \|
\delta\|_{1}^{2} \| \hat{\Sigma} - \Sigma\|_{\infty}. \label{e2}%
\end{equation}
Now we digress a bit to simplify (\ref{e2}). Note that we have the restriction
set definition
\[
\| \delta_{S_{0}^{c}} \|_{1} \le3 \sqrt{s_{0}} \| \delta_{S_{0}} \|_{2},
\]
where we add $\| \delta_{S_{0}} \|_{1}$ to both sides
\begin{align*}
\| \delta\|_{1}  &  \le3 \sqrt{s_{0}} \| \delta_{S_{0}} \|_{2} + \|
\delta_{S_{0}} \|_{1}\\
&  \le3 \sqrt{s_{0}} \| \delta_{S_{0}} \|_{2} + \sqrt{s_{0}} \| \delta_{S_{0}}
\|_{2}\\
&  = 4 \sqrt{s_{0 }} \| \delta_{S_{0}} \|_{2},
\end{align*}
where we used the norm inequality $\| \delta_{S_{0}} \|_{1} \le\sqrt{s_{0}} \|
\delta_{S_{0}} \|_{2}$ in the second inequality above. So we get
\[
\frac{ \| \delta\|_{1}^{2}}{\| \delta_{S_{0}}\|_{2}^{2}} \le16 s_{0}.
\]
Now divide (\ref{e2}) by $\| \delta_{S_{0}} \|_{2}^{2} > 0$ to have
\[
\frac{ | \delta^{\prime}\hat{\Sigma} \delta|}{ \| \delta_{S_{0}} \|_{2}^{2}}
\ge\frac{ | \delta^{\prime}\Sigma\delta| }{\| \delta_{S_{0}} \|_{2}^{2}} - 16
s_{0} \| \hat{\Sigma} - \Sigma\|_{\infty}.
\]
Minimize over $\delta$ on the both sides
\begin{equation}
\hat{\phi}_{\hat{\Sigma}}^{2} (s_{0}) \ge\phi_{\Sigma}^{2} (s_{0}) - 16 s_{0}
\| \hat{\Sigma} - \Sigma\|_{\infty}. \label{st1}%
\end{equation}
So if we can prove that with probability approaching one, $16 s_{0} \|
\hat{\Sigma} - \Sigma\|_{\infty} \le\phi_{\Sigma}^{2} (s_{0})/2$, that will
imply of $\hat{\phi}_{\hat{\Sigma}}^{2} (s_{0}) \ge\phi_{\Sigma}^{2}
(s_{0})/2$ with probability approaching one. Define $\epsilon_{n} = 16 s_{0}
t_{1}$, where
\begin{equation}
t_{1} = K [ \sqrt{\frac{ln p^{2}}{n}} + \frac{ \sqrt{E M_{2}^{2}} ln p^{2}}%
{n}] + \sqrt{\frac{ln p}{n}}. \label{pl3.9}%
\end{equation}
By (\ref{sa3})(\ref{sa4}), via Assumption \ref{a1}
\begin{align}
P [ 16 s_{0} \| \hat{\Sigma} - \Sigma\|_{\infty} > \epsilon_{n} ]  &  = P [ \|
\hat{\Sigma} - \Sigma\|_{\infty} > t_{1}]\nonumber\\
&  \le exp (- C_{1} ln p) + \frac{ K E M_{2}^{2}}{(n ln p)}\nonumber\\
&  \to0, \label{st2}%
\end{align}
where we use Assumption \ref{a2} for the probability tail converging to zero.
Also see that by Assumption \ref{a2}, $\epsilon_{n} \to0$ since $s_{0}
\sqrt{lnp/n} \to0$. So we get, with probability approaching one, $16 s_{0} \|
\hat{\Sigma} - \Sigma\|_{\infty} \le\epsilon_{n} \le\phi_{\Sigma}^{2} (
s_{0})/2$, since left side of that inequality converges to zero in
probability, and the right side is constant. Then by (\ref{st1})(\ref{st2})
\begin{align}
P [ \hat{\phi}_{\hat{\Sigma}}^{2} (s_{0}) \ge\phi_{\Sigma}^{2} (s_{0})/2]  &
\ge1 - exp (-C_{1} \kappa_{n}) - \frac{K EM_{2}^{2}}{(n \kappa_{n}%
)}\nonumber\\
&  = 1 - \frac{1}{p^{C_{1}}} - \frac{K E M_{2}^{2}}{n ln p }\nonumber\\
&  = 1 - o(1). \label{e3}%
\end{align}
\textbf{Q.E.D.}

We need the following Lemma for the exception set $\mathcal{F}^{c}:= \{ A_{1}
\cap A_{2} \}^{c}$ upper bound probability.

\begin{lemma}
\label{l0}

Under Assumptions \ref{a1}, \ref{a2}, with $\kappa_{n} = ln p$
\begin{align*}
P (\mathcal{F}^{c})  &  \le2 exp(-C_{1} \kappa_{n}) + \frac{K [ E M_{1}^{2} +
E M_{2}^{2}]}{(n \kappa_{n})}\\
&  = \frac{2}{p^{C_{1}}} + \frac{K (E M_{1}^{2} + E M_{2}^{2})}{n ln p} =
o(1).
\end{align*}

\end{lemma}

\textbf{Proof of Lemma \ref{l0}}.

Now we provide an upper bound for the probability $P (\mathcal{F}^{c})$ in our
case under Assumptions \ref{a1}, \ref{a2}, by using Lemmata \ref{l2}-\ref{l3}
\begin{align}
P (\mathcal{F}^{c})  &  = P (\mathcal{A}_{1} \cap\mathcal{A}_{2})^{c} = P
(\mathcal{A}_{1}^{c} \cup\mathcal{A}_{2}^{c}) \le P (\mathcal{A}_{1}^{c}) + P
(\mathcal{A}_{2}^{c})\nonumber\\
&  \le2 exp(-C_{1}\kappa_{n}) + \frac{K [ E M_{1}^{2} + E M_{2}^{2}]}{(n
\kappa_{n})}\nonumber\\
&  = \frac{2}{p^{C_{1}}} + \frac{ K [ E M_{1}^{2} + E M_{2}^{2}]}{n
lnp}\nonumber\\
&  \to0. \label{pt1.8a}%
\end{align}
\textbf{Q.E.D.}

\subsubsection{New Oracle Inequality Proofs}

We start with proof of Theorems \ref{t1}-\ref{t2}, where they are used as
inputs to proof of Theorem \ref{t3}. Theorems \ref{t1}-\ref{t2} consider the
new oracle inequalities.

\noindent\textbf{Proof of Theorem \ref{t1}}. We proceed in several steps.

Denote the joint event $\mathcal{F} = \{ \mathcal{A}_{1} \cap\mathcal{A}%
_{2}\}$. $\mathcal{F}^{c}$ is $\mathcal{F}$ 's complement. See that
\begin{equation}
E \| \hat{\beta} - \beta_{0} \|_{1}^{k} = E \| \hat{\beta} - \beta_{0}
\|_{1}^{k} 1_{ \{ \mathcal{F} \}} + E \| \hat{\beta} - \beta_{0} \|_{1}^{k}
1_{ \{ \mathcal{F}^{c} \}}. \label{pt1.0}%
\end{equation}
We want to form rates for the right side terms in (\ref{pt1.0}).

Step 1. Note that by Lemma \ref{l1}, the first term on the right side of
(\ref{pt1.0}) is:
\begin{equation}
E \| \hat{\beta} - \beta_{0} \|_{1}^{k} 1_{ \{ \mathcal{F} \} } = O( s_{0}^{k}
\lambda_{n}^{k}). \label{pt1.2}%
\end{equation}

Now we want to evaluate the second term on the right side of (\ref{pt1.0}).
But before that we need the following intermediate step.

Step 2. Use Nemirowski's moment inequality, Lemma 14.24 in \cite{bvdg2011},
with for all $k \ge1$, for the first inequality, and for the second inequality
by Loeve's $c_{r}$ inequality, and for the equality we use $u_{i}$ being iid,
also the definition of $\sigma^{2}:= E u_{i}^{2}$,
\begin{align*}
E \left|  \frac{\sum_{i=1}^{n} u_{i}^{2} - \sigma^{2}}{n} \right|  ^{k}  &
\le[8 ln(2)]^{k/2} E \left[  \frac{\sum_{i=1}^{n} (u_{i}^{4})}{n^{2}} \right]
^{k/2}\\
&  \le\frac{C n^{(k/2) -1 }}{n^{k}} \sum_{i=1}^{n} E u_{i}^{2 k}\\
&  = C [E u_{i}^{2k}] n^{-k/2} = O (n^{-k/2}) = o(1),
\end{align*}
by Assumption 1. Before the next result we provide the inequality,
\begin{equation}
| x + y |^{k} \le2^{k-1} ( | x |^{k} + |y|^{k}), \label{ni1}%
\end{equation}
for $k \ge1$, and $x,y$ being generic scalars, and $\sigma^{2}$ being bounded
above by Assumption 1 and using (\ref{ni1})
\begin{align}
E \left|  \frac{1}{n} \sum_{i=1}^{n} u_{i}^{2} \right|  ^{k}  &  = E \left|
\frac{1}{n} \sum_{i=1}^{n} (u_{i}^{2} - \sigma^{2}) + \sigma^{2} \right|
^{k}\nonumber\\
&  \le2^{k-1} \left[  E \left|  \frac{1}{n} \sum_{i=1}^{n} (u_{i}^{2} -
\sigma^{2}) \right|  ^{k} + (\sigma^{2})^{k}\right] \nonumber\\
&  = O (n^{-k/2}) + O(1) = O(1). \label{pt1.3}%
\end{align}

Step 3. Now we have to form another $l_{1}$ expectation bound for lasso that
will be key to the second right side term analysis in (\ref{pt1.0}). This step
3 modifies the proof of Theorem 1, supplement, p.4 of \cite{jvdg18}. We extend
their proof to non-sub-Gaussian case and show that their bound is very
conservative, and we provide a new less conservative bound. Start with the
definition of lasso.
\[
\Vert Y-X\hat{\beta}\Vert_{n}^{2}+2\lambda_{n}\Vert\hat{\beta}\Vert_{1}%
\leq\Vert Y-X\beta_{0}\Vert_{n}^{2}+2\lambda_{n}\Vert\beta_{0}\Vert_{1}.
\]
Ignore the first term and use the model $u=Y-X\beta_{0}$ to have
\[
\Vert\hat{\beta}\Vert_{1}\leq\frac{\Vert u\Vert_{n}^{2}}{2\lambda_{n}}%
+\Vert\beta_{0}\Vert_{1}.
\]
Then use triangle inequality and then the inequality above
\begin{equation}
\Vert\hat{\beta}-\beta_{0}\Vert_{1}\leq\Vert\hat{\beta}\Vert_{1}+\Vert
\beta_{0}\Vert_{1}\leq\frac{\Vert u\Vert_{n}^{2}}{2\lambda_{n}}+2\Vert
\beta_{0}\Vert_{1}. \label{pt1.3a}%
\end{equation}
Next taking the $k$ th moment of the sampling error in $l_{1}$ norm, and using
(\ref{ni1}) by taking expectations there for the second inequality below
\begin{equation}
E\Vert\hat{\beta}-\beta_{0}\Vert_{1}^{k}\leq E\left[  \frac{\Vert u\Vert
_{n}^{2}}{2\lambda_{n}}+2\Vert\beta_{0}\Vert_{1}\right]  ^{k}\leq
2^{k-1}\{E\left[  \frac{\Vert u\Vert_{n}^{2}}{2\lambda_{n}}\right]
^{k}+2\Vert\beta_{0}\Vert_{1}^{k}\} \label{pt1.4}%
\end{equation}
We use the assumption $\Vert\beta_{0}\Vert_{2}=O(1)$ to have
\begin{equation}
\Vert\beta_{0}\Vert_{1}^{k}\leq(\sqrt{s_{0}}\Vert\beta_{0}\Vert_{2}%
)^{k}=O(s_{0}^{k/2}). \label{pt1.4a}%
\end{equation}
Then use the last equation with (\ref{pt1.3}) in (\ref{pt1.4}) to have
\begin{equation}
E\left[  \frac{\Vert u\Vert_{n}^{2}}{2\lambda_{n}}\right]  ^{k}+2\Vert
\beta_{0}\Vert_{1}^{k}=O(\lambda_{n}^{-k})+O(s_{0}^{k/2})=O(\max(s_{0}%
^{k/2},\lambda_{n}^{-k})). \label{pt1.5}%
\end{equation}
Note that proof of \cite{jvdg18} use $s_{0}^{k/2}\lambda_{n}^{-k}$ but this is
very conservative upper bound since both two terms in multiplication is
diverging with $n$. But a better bound is $\max(s_{0}^{k/2},\lambda_{n}^{-k})$.

We get the rough bound for expectation using (\ref{pt1.5}) in (\ref{pt1.4})
\begin{equation}
E \| \hat{\beta} - \beta_{0} \|_{1}^{k}= O (\max(s_{0}^{k/2},\lambda_{n}%
^{-k})). \label{pt1.6}%
\end{equation}
Note that rates in (\ref{pt1.2})(\ref{pt1.6}) are different and the last rate
in this step is a rough bound which will be helpful in the next step. The rate
in (\ref{pt1.6}) is diverging to infinity.

We can simplify the rate further, since we are aiming for an asymptotic result
for incentive compatibility, with sufficiently large $n$, by Assumption 2,
$s_{0}^{k/2} \lambda_{n}^{k} = \frac{s_{0}^{k} \lambda_{n}^{k}}{s_{0}^{k/2}}
\le1$. This last inequality implies that with sufficiently large $n$, the rate
in (\ref{pt1.6}) is
\begin{equation}
E \| \hat{\beta} - \beta_{0} \|_{1}^{k}= O (\lambda_{n}^{-k}). \label{pt1.6aa}%
\end{equation}

Step 4. Rewrite the expectation using event $\mathcal{F}, \mathcal{F}^{c}$.
\begin{align}
E \| \hat{\beta} - \beta_{0} \|_{1}^{k}  &  = E \| \hat{\beta} - \beta_{0}
\|_{1}^{k} 1_{ \{ \mathcal{F} \} } + E \| \hat{\beta} - \beta_{0} \|_{1}^{k}
1_{ \{ \mathcal{F}^{c} \} }\nonumber\\
&  \le O (s_{0}^{k} \lambda_{n}^{k}) + \sqrt{ E \| \hat{\beta} - \beta_{0}
\|_{1}^{2k}} \sqrt{ E 1_{ \{ \mathcal{F}^{c} \}}}\nonumber\\
&  = O ( s_{0}^{k} \lambda_{n}^{k}) + O (\lambda_{n}^{-k}) \sqrt{ P (
\mathcal{F}^{c})} \label{pt1.6a}%
\end{align}
where we use (\ref{pt1.2}) and Cauchy-Schwartz inequality for the first
inequality, and the second equality is by (\ref{pt1.6}) with sufficiently
large $n$.



We can get the rate:\newline

%

\begin{equation}
s_{0}^{k} \lambda_{n}^{k} \ge\lambda_{n}^{-k} P (\mathcal{F}^{c})^{1/2}.
\label{pt1.8}%
\end{equation}

By (\ref{pt1.6a})(\ref{pt1.8})%

\[
E \| \hat{\beta} - \beta_{0} \|_{1}^{k} = O (s_{0}^{k} \lambda_{n}^{k}).
\]

We can simplify further (\ref{pt1.8}),
\begin{equation}
\lambda_{n} \ge P (\mathcal{F}^{c})^{1/4k}/s_{0}^{1/2}. \label{tc2}%
\end{equation}

So if $\lambda_{n} \ge P (\mathcal{F}^{c})^{1/4k}/s_{0}^{1/2}$ then
\begin{equation}
E \| \hat{\beta} - \beta_{0} \|_{1}^{k} = O (s_{0}^{k} \lambda_{n}^{k}).
\label{bias1}%
\end{equation}

The uniformity over $\mathcal{B}_{l_{0}} (s_{0})$ follows since the rates in
(\ref{pt1.2})(\ref{pt1.6})-(\ref{pt1.8}) depends on $\beta_{0}$ only by
$s_{0}$. \textbf{Q.E.D.}

Remark. Proof of Theorem 1 in \cite{jvdg18}, in their appendix, p.5, shows
that they use assumption with $P (\mathcal{F}^{c})$ bound chosen as in
(\ref{pfcsg}) below
\begin{equation}
\lambda_{n}\geq\frac{P(\mathcal{F}^{c})^{1/4k}}{s_{0}^{1/4}}, \label{r1a}%
\end{equation}
which is equivalent to the following condition as shown in p.3 of proof of
Theorem 1 in \cite{jvdg18}
\[
\tau^{2}>2kln[(\sqrt{s_{0}}\lambda_{n}^{2})^{-1}]/lnp,
\]
given that $\lambda_{n}\geq C\tau\sqrt{lnp/n}$ and $C>0,\tau>1$ with
\begin{equation}
P(\mathcal{F}^{c})\leq\frac{2}{(2p)^{\tau^{2}/2}} \label{pfcsg}%
\end{equation}
by Lemma 7 in appendix of \cite{jvdg18}. Our result and theirs are not
comparable in terms of $\lambda_{n}$ since they assume sub-Gaussian data, and
ours is more general, and their upper bound in (\ref{pfcsg}) is different than
our Lemma A.4.

\textbf{Proof of Theorem \ref{t2}}.

We start with
\begin{equation}
E \| \hat{\beta} \|_{1}^{k} = E \| \hat{\beta} \|_{1}^{k} 1_{ \{ \mathcal{F}
\} } + E \| \hat{\beta} \|_{1}^{k} 1_{ \{ \mathcal{F}^{c} \} } \le E \|
\hat{\beta} \|_{1}^{k} 1_{ \{ \mathcal{F} \} } + \sqrt{ E \| \hat{\beta}
\|_{1}^{2k}} \sqrt{ P ( \mathcal{F}^{c})}, \label{pt2.1}%
\end{equation}
by using Cauchy-Schwartz inequality. Then use triangle inequality on set
$\mathcal{F}$ and by Lemma \ref{l1}, and norm inequality to have
\begin{align*}
\| \hat{\beta} \|_{1}  &  \le\| \hat{\beta} - \beta_{0} \|_{1} + \| \beta_{0}
\|_{1}\\
&  \le\frac{24 \lambda_{n} s_{0}}{\phi_{\Sigma}^{2} (s_{0})} + \sqrt{s_{0}} \|
\beta_{0} \|_{2}\\
&  = O_{p} (\sqrt{s_{0}}),
\end{align*}
by Assumptions \ref{a1}, \ref{a2}. This last rate shows that
\begin{equation}
E \| \hat{\beta} \|_{1}^{k} 1_{ \{ \mathcal{F} \}} = O (s_{0}^{k/2}).
\label{pt2.2}%
\end{equation}

To handle the second right side term in (\ref{pt2.1}) we start with the second
inequality in (\ref{pt1.3a}) and ignore $\| \beta_{0} \|_{1}$ in the middle to
have
\[
\| \hat{\beta} \|_{1} \le\frac{ \|u \|_{n}^{2}}{2 \lambda_{n}} + \| \beta_{0}
\|_{1}.
\]
then follow (\ref{pt1.5}) to get
\begin{align}
\sqrt{ E \| \hat{\beta} \|_{1}^{2k}} P (\mathcal{F}^{c})^{1/2}  &  = O (\max(
s_{0}^{k/2}, \lambda_{n}^{-k})) P (\mathcal{F}^{c})^{1/2}\nonumber\\
&  = O (\lambda_{n}^{-k} P (\mathcal{F}^{c})^{1/2}), \label{pt2.3}%
\end{align}
and to get the second equality by Assumption \ref{a2}(ii) $(s_{0} \lambda
_{n})^{k}/s_{0}^{k/2} \le1$ since the ratio on the left converges to zero, so
this means $s_{0}^{k/2} \le\lambda_{n}^{-k}$ with sufficiently large $n$.

Now use (\ref{pt2.2}) with (\ref{pt2.3}) in (\ref{pt2.1})%

\begin{equation}
E \| \hat{\beta} \|_{1}^{k} = O (s_{0}^{k/2}) + O (\lambda_{n}^{-k} P
(\mathcal{F}^{c})^{1/2}). \label{pt2.4}%
\end{equation}


If $\lambda_{n} \ge P(\mathcal{F}^{c})^{1/2k}/s_{0}^{1/2}$ it is clear that%

\begin{equation}
s_{0}^{k/2} \ge\lambda_{n}^{-k} P(\mathcal{F}^{c})^{1/2}, \label{pt2.11}%
\end{equation}

So by (\ref{pt2.11}) in (\ref{pt2.4}) we have the desired result.
\textbf{Q.E.D.}


\textbf{Q.E.D.}

\subsubsection{Main Theorem Proof: Incentive Compatibility}

\label{app4}

\textbf{Proof of Theorem \ref{t3}}.

By Theorem \ref{t1} and \ref{t2} we can choose the larger of $\lambda_{n}$ in
those theorems, with $s_{0} \ge1$, and since it is nondecreasing with $n$,
\begin{equation}
\lambda_{n} \ge\frac{P (\mathcal{F}^{c})^{1/4k}}{s_{0}^{1/2}} \ge\frac{P
(\mathcal{F}^{c})^{1/2k}}{s_{0}^{1/2}}%
\end{equation}

\noindent Add and subtract $X_{n+1}^{\prime}\hat{\beta}$ inside the right hand
side of the incentive compatibility definition:%

\begin{align}
E [ R (X_{n+1})^{\prime}\hat{\beta} - X_{n+1}^{\prime}\beta_{0}]^{2}  &  = E [
R (X_{n+1})^{\prime}\hat{\beta} - X_{n+1}^{\prime}\hat{\beta} + X_{n+1}%
^{\prime}\hat{\beta} - X_{n+1}^{\prime}\beta_{0}]^{2}\nonumber\\
&  = E [ R (X_{n+1})^{\prime}\hat{\beta} - X_{n+1}^{\prime}\hat{\beta}]^{2} +
E[ X_{n+1}^{\prime}\hat{\beta} - X_{n+1}^{\prime}\beta_{0}]^{2}\nonumber\\
&  + E [ \hat{\beta}^{\prime}(R (X_{n+1}) - X_{n+1}) X_{n+1}^{\prime}(
\hat{\beta} - \beta_{0})]\nonumber\\
&  + E [ (\hat{\beta} - \beta_{0})^{\prime}X_{n+1} (R (X_{n+1})^{\prime
}-X_{n+1}^{\prime}) \hat{\beta}]. \label{pt3.1}%
\end{align}

\noindent Using the definition of incentive compatibility, with defining
$D_{n+1}:= R (X_{n+1}) - X_{n+1}$, we have
\begin{align}
E [ R (X_{n+1})^{\prime}\hat{\beta} - X_{n+1}^{\prime}\beta_{0}]^{2} - E[
X_{n+1}^{\prime}\hat{\beta} - X_{n+1}^{\prime}\beta_{0}]^{2}  &  = E [
\hat{\beta}^{\prime} D_{n+1} D_{n+1}^{\prime}\hat{\beta}]\label{pt3.2}\\
&  + E [ \hat{\beta}^{\prime} D_{n+1} X_{n+1}^{\prime}( \hat{\beta} -
\beta_{0})]\label{pt3.3}\\
&  + E [ (\hat{\beta} - \beta_{0})^{\prime} X_{n+1} D_{n+1}^{\prime}\hat
{\beta}]. \label{pt3.4}%
\end{align}

Now analyze (\ref{pt3.3}), the analysis of (\ref{pt3.4}) is the same and thus
omitted. See that
\begin{align}
\hat{\beta}^{\prime} D_{n+1} X_{n+1}^{\prime}(\hat{\beta} - \beta_{0})  &
\le| \hat{\beta}^{\prime} D_{n+1} X_{n+1}^{\prime}(\hat{\beta} - \beta
_{0})|\nonumber\\
&  \le|\hat{\beta}^{\prime}D_{n+1}| |X_{n+1}^{\prime}(\hat{\beta} - \beta
_{0})|\nonumber\\
&  \le\| \hat{\beta} \|_{1} \| D_{n+1} \|_{\infty} \| X_{n+1} \|_{\infty} \|
\hat{\beta} - \beta_{0} \|_{1}, \label{pt3.4a}%
\end{align}
where we use Holder's inequality. Then
\begin{align}
E [ \hat{\beta}^{\prime} D_{n+1} X_{n+1}^{\prime}( \hat{\beta} - \beta_{0})]
&  \le\| D_{n+1} \|_{\infty} \| X_{n+1} \|_{\infty} \| E \left[  \| \hat
{\beta} \|_{1} \| \hat{\beta} - \beta_{0} \|_{1} \right] \label{pt3.5}\\
&  \le\left[  \| D_{n+1} \|_{\infty}\right]  \left[  \| X_{n+1} \|_{\infty
}\right]  \left[  E \| \hat{\beta}\|_{1}^{2}\right]  ^{1/2} \left[  E \|
\hat{\beta} - \beta_{0} \|_{1}^{2}\right]  ^{1/2}\\
&  = [ M_{4}] [ M_{3}] \left[  E \| \hat{\beta}\|_{1}^{2}\right]  ^{1/2}
\left[  E \| \hat{\beta} - \beta_{0} \|_{1}^{2}\right]  ^{1/2} \label{pt3.6}%
\end{align}
where we apply (\ref{pt3.4a}) for the first inequality and Holder's Inequality
in the second inequality above, and the last equality comes from $M_{3},
M_{4}$ definitions. Then we apply Theorems \ref{t1}-\ref{t2} with $k=2$. We
assume $\lambda_{n} \ge P ( \mathcal{F}^{c})^{1/8}/s_{0}^{1/2}$ and if
\begin{equation}
s_{0}^{3/2} \sqrt{\frac{lnp}{n}} [ M_{3}] [ M_{4}] \to0, \label{a3}%
\end{equation}
we see that (\ref{pt3.6}) goes to zero, by Theorems \ref{t1}-\ref{t2}, and
$\lambda_{n} = O (\sqrt{\frac{lnp}{n}})$.



So looking at incentive compatibility definition and (\ref{pt3.2}%
)-(\ref{pt3.4})
\begin{equation}
E [ R (X_{n+1})^{\prime}\hat{\beta} - X_{n+1}^{\prime}\beta_{0}]^{2} - E[
X_{n+1}^{\prime}\hat{\beta} - X_{n+1}^{\prime}\beta_{0}]^{2} = E [ \hat{\beta
}^{\prime}D_{n+1} D_{n+1}^{\prime}\hat{\beta}] + o(1), \label{pt3.9}%
\end{equation}
where the first right side term in (\ref{pt3.9}) is nonnegative and the other
terms are negligible in large samples by (\ref{a3}).


The uniformity over $\mathcal{B}_{l_{0}} (s_{0})$ goes through since Theorems
\ref{t1}, \ref{t2} depend on $\beta_{0}$ only through $s_{0}$, and they are
the main ingredient in the proof.

\textbf{Q.E.D.}

\setcounter{equation}{0}\setcounter{lemma}{0}\renewcommand{\theequation}{B.\arabic{equation}}\renewcommand{\thelemma}{B.\arabic{lemma}}

\section{Appendix B}

Here we consider results when $p \le n$, and relaxing Assumption \ref{a2}(iii).

\subsection{When $p \le n$}

There are minor modifications in the proofs compared to $p>n$. We consider
them here. One major change is since $p\le n $, we set $\kappa_{n} = ln n$.
Change Assumption 2(ii) so that $s_{0} \sqrt{ln/n} \to0$.

We provide the maximal inequality here. Now take the case of $p\le n$, and
combine (\ref{sa2}) with (\ref{sa3}) to have with $\kappa_{n} = ln n$ in that
case
\begin{align}
P ( \max_{1 \le j \le p} | \hat{\mu}_{j} - \mu_{j} |  &  \ge2K [ \frac
{\sqrt{lnp}}{\sqrt{n}} + \frac{ (E M_{F}^{2})^{1/2} ln p}{n}] + \frac
{\sqrt{lnn}}{\sqrt{n}})\nonumber\\
&  \le\frac{1}{n^{C_{1}}} + \frac{ E M_{F}^{2}}{n (ln n)} = o(1), \label{sa6}%
\end{align}
by Assumptions A1-A.2. To see this point
\begin{equation}
\frac{ E M_{F}^{2}}{n ln n} = \left[  \left(  \frac{(E M_{F}^{2})^{1/2}
\sqrt{lnp}}{\sqrt{n}} \right)  \frac{1}{\sqrt{ln n} \sqrt{lnp} } \right]  ^{2}
= o(1). \label{aux1}%
\end{equation}
This shows also that
\begin{equation}
\max_{1 \le j \le p } | \hat{\mu}_{j} - \mu_{j} | = O_{p} ( \sqrt{lnn}%
/\sqrt{n}). \label{sa7}%
\end{equation}

Lemma \ref{l1} will be the same. Lemma \ref{l2}(i) lower bound probability has
$\kappa_{n} = ln n$ now. Lemma \ref{l2}(ii) is the same. Lemma \ref{l2}(iii)
will change to $\lambda_{n} = O (\sqrt{ln n}/\sqrt{n}).$ Lemma \ref{l3} use
$\kappa_{n} = ln n$, so (\ref{pl3.9}) becomes
\[
t_{1} = K [ \frac{\sqrt{ln p^{2}}}{\sqrt{n}} + \frac{ \sqrt{E M_{2}^{2}} ln
p^{2}}{n}] + \frac{\sqrt{ln n}}{n}.
\]

Lemma \ref{l0} is the same with $\kappa_{n} = ln n$.

Given these results, the proof of Theorem \ref{t1} is the same with
$\lambda_{n} = O (\sqrt{\frac{ln n}{n}})$. Theorem 2 does not change. Theorem
\ref{t3} condition will be changing to
\[
s_{0}^{3/2} \sqrt{\frac{ln n}{n}} [ M_{3}] [ M_{4}] \to0,
\]

\subsection{Relaxing Assumption \ref{a2}(iii)}

In this subsection we relax Assumption \ref{a2}(iii) from $\| \beta_{0}\|_{2}
= O (1)$ to $\| \beta_{0} \|_{2} = O (\sqrt{s_{0}})$ and we explain the logic
and meaning of this new assumption.

\textbf{Assumption \ref{a2}(iv)}.\textit{%
\[
\| \beta_{0} \|_{2} = O (\sqrt{s_{0}}).
\]
}

Assumption \ref{a2}(iii) which is suggested by \cite{jvdg18} and simplifies
their paper in semiparametric efficient estimators. Our Assumption 2(iv) here
generalizes that assumption and in the case of $s_{0}$ being constant becomes
Assumption 2(iii). The implication of Assumption 2(iv) is that all nonzero
coefficients can be constant and none of them has to be local to zero.
\[
\| \beta_{0} \|_{2} = \sqrt{ \sum_{j=1}^{p} \beta_{0,j}^{2}} = \sqrt{ \sum_{j
\in S_{0}} \beta_{0,j}^{2}} = O (\sqrt{s_{0}}).
\]
In terms of Section 2 discussion after Assumption 2, this implies $S_{0} =
F_{1}$, and $F_{2}$ is an empty set. So Assumption 2(iv) can simultaneously
allow $s_{0}$ increasing with $n$, and all large nonzero coefficients in
$S_{0}$. Previously in Assumption 2(iii), there can be only a fixed number of
large coefficients, and increasing ($s_{0}-f_{1}$) number of local to zero
(small) coefficients.

We proceed in a way that we only change the proofs in Appendix A, when
necessary. All lemmata in Appendix A goes through, there is no usage of
Assumption 2(iii) there. The first change comes in step 3 of Theorem 1 proof.
First (\ref{pt1.4a}) changes to $\|\beta_{0} \|_{1}^{k} = O (s_{0}^{k})$ under
Assumption 2(iv) instead of Assumption 2(iii). Then (\ref{pt1.5}) becomes
\begin{equation}
E \left[  \frac{\| u \|_{n}^{2}}{2 \lambda_{n}} \right]  ^{k} + 2 \| \beta_{0}
\|_{1}^{k} = O (\max(s_{0}^{k}, \lambda_{n}^{-k})). \label{b1}%
\end{equation}

Then (\ref{pt1.6a}) changes to following
\begin{align}
E \| \hat{\beta} - \beta_{0} \|_{1}^{k}  & = O (s_{0}^{k} \lambda_{n}^{k})+ O
(\max(s_{0}^{k}, \lambda_{n}^{-k})\sqrt{P (\mathcal{F}^{c})}\nonumber\\
&  = O (s_{0}^{k} \lambda_{n}^{k})+ O(\lambda_{n}^{-k}\sqrt{P (\mathcal{F}%
^{c}))} ,\label{b2}%
\end{align}
where we use Assumption 2 with $s_{0}^{k} \lambda_{n}^{k} \le1 $ and
sufficiently large $n$ to show the last equality. Instead of (\ref{pt1.8}) we
have the following conditions, to establish the rate for the oracle inequality
(i.e. mean $l_{1}$ norm bound to $k$ th order)
\begin{equation}
s_{0}^{k} \lambda_{n}^{k} \ge\lambda_{n}^{-k} P (\mathcal{F}^{c})^{1/2}.
\label{b4}%
\end{equation}
Using (\ref{b2})-(\ref{b4})
\begin{equation}
E \| \hat{\beta} - \beta_{0} \|_{1}^{k} = O (s_{0}^{k} \lambda_{n}^{k}).
\label{b5}%
\end{equation}
The condition (\ref{b4}) can be written as
\begin{equation}
\lambda_{n} \ge P (\mathcal{F}^{c})^{1/4k}/s_{0}^{1/2}, \label{b6}%
\end{equation}
where the tuning parameter choice under Assumption 2(iv) which is (\ref{b6})
is the same. The discussion after this in step 4 is the same, given Assumption
\ref{a2}(i)-(ii). So we have the following result:

\textbf{Corollary B.1}. \textit{Under Assumptions 1, 2(i)(ii)(iv), with
sufficiently large n
\[
\lambda_{n} \ge P (\mathcal{F}^{c})^{1/4k}/s_{0}^{1/2}.
\]
we have
\[
[E \| \hat{\beta} - \beta_{0} \|_{1}^{k}]^{1/k} = O (s_{0} \lambda_{n}).
\]
The result is also uniform over $l_{0}$ ball $\mathcal{B}_{l_{0}}$ }

Now we modify the proof of Theorem \ref{t2}. In that respect, by Assumption
2(iv) the rate after (\ref{pt2.1}) becomes
\begin{equation}
\| \hat{\beta} \|_{1} = O_{p} (s_{0}). \label{b7}%
\end{equation}
Then (\ref{pt2.4}) changes to
\begin{equation}
E \| \hat{\beta} \|_{1}^{k} = O (s_{0}^{k}) + O (\lambda_{n}^{-k} P
(\mathcal{F}^{c})^{1/2}). \label{b8}%
\end{equation}
We can show that
\begin{equation}
s_{0}^{k} \ge\lambda_{n}^{-k} P (\mathcal{F}^{c})^{1/2}, \label{b9}%
\end{equation}
if we have
\begin{equation}
\lambda_{n} \ge P (\mathcal{F}^{c})^{1/2k}/s_{0}. \label{b10}%
\end{equation}
Then given (\ref{b10}), using (\ref{b9}) in (\ref{b8}) we have
\[
E \| \hat{\beta} \|_{1}^{k} = O (s_{0}^{k}).
\]
So we established the following Corollary to Theorem \ref{t2}. The result is
different from Theorem \ref{t2} and the k th moment of $l_{1}$ error grows
faster here in Corollary B.2 if $s_{0}$ increases with $n$. So relaxed
assumption comes with a cost that will affect main incentive compatibility condition.

\textbf{Corollary B.2}. \textit{Under Assumptions 1, 2(i)(ii)(iv), with
sufficiently large n
\[
\lambda_{n} \ge P (\mathcal{F}^{c})^{1/2k}/s_{0}.
\]
we have
\[
[E \| \hat{\beta} \|_{1}^{k}]^{1/k} = O (s_{0}).
\]
The result is also uniform over $l_{0}$ ball $\mathcal{B}_{l_{0}}$}

Now we follow the proof of Theorem \ref{t3} and substitute Assumption 2(iv)
instead of Assumption 2(iii). Note that our $\lambda_{n}$ choice must choose
the maximum of the ones in Corollary B.1 and B.2. Clearly Corollary B.1 tuning
parameter is larger than the one in Corollary B.2. The only place we have to
change there is (\ref{a3}). Given
\[
\lambda_{n} \ge max \left(  \frac{P (\mathcal{F}^{c})^{1/4}}{s_{0}}, \frac{P
(\mathcal{F}^{c})^{1/8}}{s_{0}^{1/2}}\right)  = \frac{P (\mathcal{F}%
^{c})^{1/8}}{s_{0}^{1/2}},
\]
since $s_{0} \ge1$ we need
\[
s_{0}^{2} \sqrt{\frac{ln p}{n}} [ M_{3}] [ M_{4}] \to0,
\]
to have Incentive Compatibility in large samples. So we have the following
counterpart to Theorem 3.

\textbf{Corollary B.3}.\textit{Under Assumptions 1, 2(i)(ii)(iv) and with
sufficiently large n
\[
\lambda_{n} \ge\frac{P (\mathcal{F}^{c})^{1/8}}{s_{0}^{1/2}}
\]
and
\[
s_{0}^{2} \sqrt{\frac{ln p}{n}} [ M_{3}] [ M_{4}] \to0,
\]
lasso is Incentive Compatible. The result is also uniform over $l_{0}$ ball
$\mathcal{B}_{l_{0}}$.}

Clearly, there is a difference between Theorem \ref{t3} and Corollary B.3
here. Incentive compatibility of lasso is more difficult to achieve, due to
sparsity, $s_{0}$, having exponent of 2 here instead of 3/2 in Theorem 3.

\setcounter{equation}{0}\setcounter{lemma}{0}\renewcommand{\theequation}{C.\arabic{equation}}\renewcommand{\thelemma}{C.\arabic{lemma}}

\section{Appendix C}

This section provides the proofs for conservative lasso IC, which is explained
in Section \ref{iccl}. Let wpa1 denote with probability approaching one.
First, we start with $l_{\infty}$ bound for Lasso estimator. This bound is
needed for Conservative Lasso for the proofs of moment bounds.

\begin{lemma}
\label{lc1}

(i). Under Assumption 1, and on $\mathcal{A}_{1} \cap\mathcal{A}_{2}$
\[
\| \hat{\beta} - \beta_{0} \|_{\infty} \le\| \Theta\|_{l_{\infty}} [
\frac{\lambda_{n}}{2} + t_{1} \frac{24 \lambda_{n} s_{0}}{\phi_{\Sigma}^{2}
(s_{0})} + \lambda_{n}],
\]
with the definition
\[
\lambda_{prec}:= \| \Theta\|_{l_{\infty}} [ \frac{\lambda_{n}}{2} + t_{1}
\frac{24 \lambda_{n} s_{0}}{\phi_{\Sigma}^{2} (s_{0})} + \lambda_{n}].
\]

$t_{1}$ is defined in (\ref{pl3.9}).

(ii). With added Assumption 2 to (ii), and assuming $\| \Theta\|_{l_{\infty}}
= O(s_{1})$, $\lambda_{prec} = O (s_{1} \lambda_{n})$.

(iii). The result in (i) holds wpa1 (i.e. with probability approaching one)
with Assumptions 1-3, and
\[
\| \hat{\beta} - \beta_{0} \|_{\infty} = O_{p} (\lambda_{prec}) = O_{p} (s_{1}
\lambda_{n}) = o_{p} (1).
\]

\end{lemma}

Remark. Result (iii) holds without the need to be in $\mathcal{A}_{1}
\cap\mathcal{A}_{2}$. This is Lemma A.7 of \cite{ck18}, where we prove it
under Assumptions 1-2 which are weaker moment conditions than the one in
\cite{ck18} due to usage of new maximal inequalities in Section A.2. Also $\|
\Theta\|_{l_{\infty}} = O(s_{1})$ allows the row sums of precision matrix to
be diverging with $n$. Hence we relax the restrictive assumption of constant
maximum row sum of the precision matrix in \cite{ck18} as well as the one in
\cite{vdg16}.

\textbf{Proof of Lemma \ref{lc1}}.

(i). By Lemma 2.5.1 of \cite{vdg2014} or (A.25) of \cite{ck18}
\[
\| \hat{\beta} - \beta_{0} \|_{\infty} \le\| \Theta\|_{\l _{\infty}} \left[
\| \frac{X^{\prime}u}{n}\|_{\infty} + \| \hat{\Sigma} - \Sigma\|_{\infty}
\|\hat{\beta} - \beta_{0} \|_{1} + \lambda_{n} \right]  .
\]
Now on $\mathcal{A}_{1} \cap\mathcal{A}_{2}$ with Lemma A.1 and (A.19)-(A.20)
\[
\| \hat{\beta} - \beta_{0} \|_{\infty} \le\| \Theta\|_{\l _{\infty}} [
\frac{\lambda_{n}}{2} + t_{1} \frac{24 \lambda_{n} s_{0}}{\phi_{\Sigma}^{2}
(s_{0})} + \lambda_{n}].
\]

(ii). So we define
\[
\lambda_{prec}:= \| \Theta\|_{\l _{\infty}} \lambda_{n} [ \frac{3}{2} +
\frac{24 t_{1} s_{0}}{\phi_{\Sigma}^{2} (s_{0})}],
\]
with Assumption 2 added, and since $t_{1} = O ( \sqrt{lnp/n})$ and so $s_{0}
\sqrt{lnp/n} = o(1)$ by Assumption 2(ii) we get $\lambda_{prec}= O (s_{1}
\lambda_{n})$, given that $\| \Theta\|_{l_{\infty}} = O(s_{1})$, and
$\phi_{\Sigma}^{2} (s_{0}) \ge c > 0$, for $c>0$ is a positive constant.

(iii). This is true by Lemma A.2-Lemma A.3 given (ii) and Assumption
3.\textbf{Q.E.D.}

We have the following $l_{1}$ norm result, which is Lemma A.1 in \cite{ck18}.
Their assumptions are slightly stronger, with our new Lemma A.2-A.3 for the
sets $\mathcal{A}_{1}, \mathcal{A}_{2}$ we can prove under our Assumptions
1-2, part (ii) of Lemma C.2. Part (i) below is from their paper.

\begin{lemma}
\label{lc2} Let $0 < a_{n} \le1$, where $a_{n}$ is a deterministic-positive
sequence in $n$, then

(i). on the set $\mathcal{A}_{1} \cap\mathcal{A}_{2}$
\[
\| \hat{\beta}_{w} - \beta_{0} \|_{1} \le4 (a_{n} +1) (2 a_{n}+1)
\frac{\lambda_{n} s_{0}}{\phi_{\Sigma}^{2} (s_{0})}.
\]

(ii). with Assumptions 1-2
\[
\| \hat{\beta}_{w} - \beta_{0} \|_{1} = O_{p} ( \lambda_{n} s_{0}).
\]

\end{lemma}

We start with the proof of moments for conservative lasso's moments. This is
extending Theorem 1 to a more general weighted penalty.

\textbf{Proof of Theorem \ref{t4}}. The proof will mirror proof of Theorem 1
above. We show the places that will differ.

Step 1. Using Lemma \ref{lc2} above
\begin{equation}
E \| \hat{\beta}_{w} - \beta_{0} \|_{1}^{k} 1_{ \{ \mathcal{F}^{c} \} } = O
(s_{0}^{k} \lambda_{n}^{k}). \label{ptc1-1}%
\end{equation}

Step 2. This is exactly the same in Step 2, Theorem 1. It only involved error
terms not the penalty. (A.25)-(A.26) are valid here as well.

Step 3. This step is a major extension of step 3 for Theorem 1, and extends
the lasso penalty and its moments to a more general-data dependent
weighted-conservative lasso. Using the definition for conservative lasso
\[
\| Y - X \hat{\beta}_{w} \|_{n}^{2} + 2 \lambda_{n} \sum_{j=1}^{p} \hat{w}_{j}
| \hat{\beta}_{w,j}| \le\| Y - X \beta_{0} \|_{n}^{2} + 2 \lambda_{n}
\sum_{j=1}^{p} \hat{w}_{j} | \beta_{0,j}|.
\]
Ignoring the first term since its nonnegative and $u:= Y - X \beta_{0}$
\begin{equation}
\sum_{j=1}^{p} \hat{w}_{j} | \hat{\beta}_{w,j} | \le\frac{ \| u \|_{n}^{2}}{2
\lambda_{n}} + \sum_{j=1}^{p} \hat{w}_{j} | \beta_{0,j} |. \label{ptc1-2}%
\end{equation}
Thus since $\max_{1 \le j \le p} \hat{w}_{j} \le1$, and define $\hat{w}%
_{min}:= \min_{1 \le j \le p} \hat{w}_{j}$ we can rewrite (\ref{ptc1-2})
\begin{equation}
\| \hat{\beta}_{w} \|_{1} \le\frac{\|u\|_{n}^{2}}{2 \lambda_{n} \hat{w}_{min}
} + \frac{ \| \beta_{0} \|_{1}}{\hat{w}_{min}}. \label{ptc1-3}%
\end{equation}
Use triangle inequality
\[
\| \hat{\beta}_{w} - \beta_{0} \|_{1} \le\| \hat{\beta}_{w} \|_{1} + \|
\beta_{0} \|_{1}.
\]
Then take expectations above and use (\ref{ptc1-3}) and (A.25)
\begin{align}
E \| \hat{\beta}_{w} - \beta_{0} \|_{1}^{k}  &  \le2^{k-1} \left\{  E \left[
\frac{ \| u \|_{n}^{2}}{2 \lambda_{n} \hat{w}_{min}} \right]  ^{k} + E \left[
\frac{2 \| \beta_{0} \|_{1}}{\hat{w}_{min}} \right]  ^{k} \right\} \nonumber\\
&  = 2^{k-1} \left\{  \frac{1}{(2 \lambda_{n})^{k}} E \left[  \frac{ \| u
\|_{n}^{2}}{ \hat{w}_{min}} \right]  ^{k} + [2 \| \beta_{0} \|_{1}]^{k} E
\left[  \frac{1}{\hat{w}_{min}} \right]  ^{k} \right\} \nonumber\\
&  \le2^{k-1} \left\{  \frac{1}{(2 \lambda_{n})^{k}} [E \| u \|_{n}%
^{4k}]^{1/2} [E (\hat{w}_{min})^{-2k}]^{1/2} + [2 \| \beta_{0} \|_{1}]^{k} E
\left[  \frac{1}{\hat{w}_{min}} \right]  ^{k} \right\}  , \label{ptc1-4}%
\end{align}
where we use Cauchy-Schwartz inequality for the first term on the right side
to get the last inequality. We consider the term $\hat{w}_{min}^{-1}$ in
(\ref{ptc1-4})
\[
\hat{w}_{min}^{-1} = \frac{\max_{1 \le j \le p} | \hat{\beta}_{j} |\cup
\lambda_{prec}}{\lambda_{prec}}.
\]

If $\max_{ 1 \le j \le p } | \hat{\beta}_{j} | \le\lambda_{prec}$ then
$\hat{w}_{min}^{-1}=1$. With that estimated minimum weight, the proofs of
Theorem 1 can go forward, but unfortunately since estimated minimum weight can
take another value and make the problem and the proofs more complicated. Now
we show this issue. If $\max_{1 \le j \le p} | \hat{\beta}_{j} | >
\lambda_{prec}$ then
\[
\hat{w}_{min}^{-1} = \frac{\max_{1 \le j \le p} | \hat{\beta}_{j} |}%
{\lambda_{prec}}.
\]
Then with with probability approaching one, given Lemma C.1
\begin{align}
\hat{w}_{min}^{-1}  & \le \frac{ \max_{1 \le j \le p} | \hat{\beta}_{j} -
\beta_{0,j}| + \max_{1 \le j \le p} | \beta_{0,j}|}{\lambda_{prec}}\\
&  \le \frac{\lambda_{prec} + C }{\lambda_{prec}} = 1 + \frac{C}%
{\lambda_{prec}}.
\end{align}
By Assumption 3 we know $\lambda_{prec} = o(1)$, via Lemma C.1
\[
\hat{w}_{min}^{-1} = O_{p} (\lambda_{prec}^{-1}).
\]

Regardless of whether $|\hat{\beta}_{j} |$ is larger than or equal to or less
than $\lambda_{prec}$ we have
\[
\frac{1}{\hat{w}_{min}} = O_{p} (s_{1}^{-1} \lambda_{n}^{-1}),
\]
since its diverging in $n$ when $|\hat{\beta}_{j} | > \lambda_{prec}$, and one
otherwise. So
\[
E \left(  \frac{1}{\hat{w}_{min}^{k}} \right)  = O (s_{1}^{-k} \lambda
_{n}^{-k}).
\]
Also
\[
\left[  E \left(  \frac{1}{\hat{w}_{min}^{2k}} \right)  \right]  ^{1/2} = O
(s_{1}^{-k} \lambda_{n}^{-k}),
\]
as well. With these two rates and by (A.26)(A.29) in (\ref{ptc1-4})%

\begin{align}
&  \frac{1}{(2 \lambda_{n})^{k}} [E \| u \|_{n}^{4k}]^{1/2} [E (\hat{w}%
_{min})^{-2k}]^{1/2} + [2 \| \beta_{0} \|_{1}]^{k} E \left[  \frac{1}{\hat
{w}_{min}}\right]  ^{k}\nonumber\\
&  = O (\lambda_{n}^{-k}) O (1) O (\lambda_{prec}^{-k}) + O (s_{0}^{k/2}) O
(\lambda_{prec}^{-k})\nonumber\\
&  = O (\lambda_{n}^{-k}) O (s_{1}^{-k} \lambda_{n}^{-k}) + O (s_{0}^{k/2} ) O
(s_{1}^{-k} \lambda_{n}^{-k})\label{rate}\\
&  = O (s_{1}^{-k} \lambda_{n}^{-2k}), \label{ptc1-5}%
\end{align}
where we use $s_{0}^{k/2} \lambda_{n}^{k} = (s_{0} \lambda_{n})^{k}%
/s_{0}^{k/2} \le1$ by Assumption 2 with sufficiently large $n$, so to get last
equality, the first rate dominated in (\ref{rate}).

So using (\ref{ptc1-5}) in (\ref{ptc1-4})
\[
E \| \hat{\beta}_{w} - \beta_{0} \|_{1}^{k} = O (s_{1}^{-k} \lambda_{n}%
^{-2k}).
\]
This is a key finding and entirely new, shows that general weight function in
conservative lasso made the error larger compared with lasso ,where weights
are all one in lasso, since we have an extra $s_{1}^{-k} \lambda_{n}^{-k}$
term extra compared with lasso in (\ref{pt1.6aa}). Extending lasso to general
weights as conservative lasso made the moment estimation worse due to weights
being very small. Conservative lasso came with better selection properties
then lasso but here it lacks in estimating moments.

Step 4. Now merge the rates in (\ref{ptc1-1})(\ref{ptc1-5})
\begin{align}
E \| \hat{\beta}_{w} - \beta_{0} \|_{1}^{k}  &  = E \| \hat{\beta}_{w} -
\beta_{0} \|_{1}^{k} 1_{ \{ \mathcal{F} \}} + E \| \hat{\beta}_{w} - \beta_{0}
\|_{1}^{k} 1_{ \{ \mathcal{F}^{c} \}}\nonumber\\
&  \le O (s_{0}^{k} \lambda_{n}^{k}) + \sqrt{ E \| \hat{\beta}_{w} - \beta_{0}
\|_{1}^{2k}} \sqrt{ P (\mathcal{F}^{c} )}\nonumber\\
&  = O (s_{0}^{k} \lambda_{n}^{k}) + O (s_{1}^{-k} \lambda_{n}^{-2k}) P
(\mathcal{F}^{c})^{1/2}. \label{ptc1-6}%
\end{align}
To establish a rate
\begin{equation}
s_{0}^{k} \lambda_{n}^{k} \ge\lambda_{n}^{-2k} s_{1}^{-k} P (\mathcal{F}%
^{c})^{1/2},\label{ptc1-7a}%
\end{equation}
which (\ref{ptc1-7a}) is implied by
\[
\lambda_{n} \ge\frac{P (\mathcal{F}^{c})^{1/6k}}{s_{0}^{1/3} s_{1}^{1/3}},
\]
This shows
\[
E \| \hat{\beta}_{w} - \beta_{0} \|_{1}^{k} = O (s_{0}^{k} \lambda_{n}^{k}).
\]
\textbf{Q.E.D}

\textbf{Proof of Theorem \ref{t5}}. By (A.43)
\begin{equation}
E \| \hat{\beta}_{w} \|_{1}^{k} \le E \| \hat{\beta}_{w} \|_{1}^{k} 1_{ \{
\mathcal{F} \}} + \sqrt{ E \| \hat{\beta}_{w} \|_{1}^{2k}} \sqrt{ P
(\mathcal{F}^{c})}. \label{ptc2-1}%
\end{equation}
By Lemma \ref{lc2}, and (A.29)
\begin{align}
\| \hat{\beta}_{w} \|_{1}  &  \le\| \hat{\beta}_{w} - \beta_{0} \|_{1} + \|
\beta_{0} \|_{1}\nonumber\\
&  = O_{p} ( \lambda_{n} s_{0}) + O (\sqrt{s_{0}}) = O_{p} ( \sqrt{s_{0}}),
\label{ptc2-2}%
\end{align}
and the last equality is by Assumption 2, since $s_{0} \lambda_{n} \to0$. So
\begin{equation}
E \| \hat{\beta}_{w} \|_{1}^{k} 1_{ \{ \mathcal{F} \}} = O (s_{0}^{k/2}).
\label{ptc2-2a}%
\end{equation}
Then to handle the second term on the right side in (\ref{ptc2-1}), use (C.3)
\[
\| \hat{\beta}_{w} \|_{1} \le\frac{ \|u \|_{n}^{2}}{2 \lambda_{n} \hat
{w}_{min}} + \frac{ \| \beta_{0} \|_{1}}{\hat{w}_{min}}.
\]
Next repeat exactly (\ref{ptc1-3})-(\ref{ptc1-5}) to have
\begin{equation}
\sqrt{ E \| \hat{\beta}_{w} \|_{1}^{2k}} \sqrt{P (\mathcal{F}^{c})} = O
(\lambda_{n}^{-2k} s_{1}^{-k}) \sqrt{P (\mathcal{F}^{c})}, \label{ptc2-3}%
\end{equation}
since $(s_{0} \lambda_{n})^{k}/s_{0}^{k/2} \le1$ for sufficiently large n. Use
(\ref{ptc2-2a})(\ref{ptc2-3}) in (\ref{ptc2-1}) to have
\begin{equation}
E \| \hat{\beta}_{w} \|_{1}^{k} = O (s_{0}^{k/2}) + O (\lambda_{n}^{-2k}
s_{1}^{-k} P (\mathcal{F}^{c})^{1/2}). \label{ptc2-3a}%
\end{equation}
Note that compared to lasso, our second rate is different by $\lambda_{n}^{-k}
s_{1}^{k}$, this is due to usage of weights, namely minimum weight estimate
being at rate of $\lambda_{n}$, and we use inverse of that estimate in the
bounds. To get a rate for $k$ th moment of conservative lasso (in $l_{1}$
norm) with
\[
\lambda_{n} \ge\frac{ P(\mathcal{F}^{c} )^{1/4k}}{s_{0}^{1/4} s_{1}^{1/2}},
\]
first rate in (\ref{ptc2-3a}) above dominates the second one, which gets us
\[
E \| \hat{\beta}_{w} \|_{1}^{k} = O (s_{0}^{k/2}).
\]
\textbf{Q.E.D.}

\textbf{Proof of Theorem 6}. Given Theorems 4-5 proof here follows exactly
from the proof of Theorem 3. But the lower bound for $\lambda_{n}$ is:
($k=2$)
\[
\lambda_{n} \ge max \left(  \frac{P ( \mathcal{F}^{c})^{1/8}}{s_{0}^{1/4}
s_{1}^{1/2}}, \frac{P (\mathcal{F}^{c})^{1/12}}{s_{0}^{1/3} s_{1}^{1/3}}
\right)
\]
.\textbf{Q.E.D.}

\bibliographystyle{chicagoa}
\bibliography{referic}

\end{document}